\documentclass{article}

\usepackage{arxiv}

\usepackage[utf8]{inputenc} % allow utf-8 input
\usepackage[T1]{fontenc}    % use 8-bit T1 fonts
\usepackage{hyperref}       % hyperlinks
\usepackage{url}            % simple URL typesetting
\usepackage{booktabs}       % professional-quality tables
\usepackage{amsfonts}       % blackboard math symbols
\usepackage{nicefrac}       % compact symbols for 1/2, etc.
\usepackage{microtype}      % microtypography
\usepackage{lipsum}

\usepackage{relsize}
\usepackage{verbatim}
\usepackage{graphicx}% Include figure files
\graphicspath{{Figures/}}
\usepackage{dcolumn}% Align table columns on decimal point
\usepackage{bm}% bold math
\usepackage{subcaption}
\usepackage{float}

\usepackage{amsmath,amssymb,amsfonts}
\usepackage{graphics}
\usepackage{color}
\usepackage{xcolor}
\definecolor{rev1}{rgb}{0,0,0}

\usepackage{cite}

\usepackage[ruled,vlined]{algorithm2e}
\usepackage[capitalise]{cleveref}

\newcommand{\bu}{\boldsymbol{u}}

\newcommand{\bx}{\boldsymbol{x}}

\newcommand{\bff}{\boldsymbol{f}}
\newcommand{\bFF}{F}%{\boldsymbol{F}}

\newcommand{\bGG}{G}%{\boldsymbol{G}}

\newcommand{\bc}{\boldsymbol{c}}
\newcommand{\ba}{\boldsymbol{a}}
\newcommand{\bb}{\boldsymbol{b}}

\newcommand{\bX}{\boldsymbol{X}}

\newcommand{\re}{\text{Re}}

\newcommand{\Bh}{\mathbf{h}}
\newcommand{\Bg}{\mathbf{g}}

\newcommand{\Bb}{\mathbf{b}}
\newcommand{\Bs}{\mathbf{s}}
\newcommand{\BW}{\mathbf{W}}

\newcommand{\bz}{\boldsymbol{z}}

\title{Physics Guided Machine Learning for Variational Multiscale Reduced Order Modeling}

\author{
 Shady E. Ahmed \\
  School of Mechanical \& Aerospace Engineering,\\
  Oklahoma State University,\\
  Stillwater, OK 74078, USA.\\
  \texttt{shady.ahmed@okstate.edu}
\And
 Omer San \\
  School of Mechanical \& Aerospace Engineering,\\
  Oklahoma State University,\\
  Stillwater, OK 74078, USA.\\
  \texttt{osan@okstate.edu} %\\
  \And
 Adil Rasheed\\
  Department of Engineering Cybernetics,\\
  Norwegian University of Science and Technology,\\
  N-7465, Trondheim, Norway.\\
  Department of Mathematics and Cybernetics,\\
  SINTEF Digital,\\ 7034 Trondheim, Norway. \\
\texttt{adil.rasheed@ntnu.no}
\And
 Traian Iliescu \\
  Department of Mathematics,\\
  Virginia Tech,\\
  Blacksburg, VA 24061, USA.\\
  \texttt{iliescu@vt.edu} %\\
  \And
 Alessandro Veneziani \\
  Department of Mathematics, \\
  Department of Computer Science,\\
  Emory University,\\
  Atlanta, GA 30322, USA.\\
  \texttt{avenez2@emory.edu} %\\
}

\begin{document}
\maketitle

\begin{abstract}
We propose a new physics guided machine learning (PGML) paradigm that leverages the variational multiscale (VMS) framework and available data to dramatically increase the accuracy of reduced order models (ROMs) at a modest computational cost. The hierarchical structure of the ROM basis and the VMS framework enable a natural separation of the resolved and unresolved ROM spatial scales. Modern PGML algorithms are used to construct novel models for the interaction among the resolved and unresolved ROM scales. Specifically, the new framework builds ROM operators that are closest to the true interaction terms in the VMS framework. Finally, machine learning is used to reduce the projection error and further increase the ROM accuracy. Our numerical experiments for a two-dimensional vorticity transport problem show that the novel PGML-VMS-ROM paradigm maintains the low computational cost of current ROMs, while significantly increasing the ROM accuracy.
\end{abstract}

 %Use showkeys class option if keyword

%%=======================================%%
\section{Introduction} \label{sec:intro}
%%=======================================%%
The behavior of physical systems can be generally described by 
physical principles (e.g., conservation of mass, momentum, and energy)
together with constitutive laws. The resulting models 
are often mathematically formulated as partial differential equations (PDEs) (e.g., the Navier-Stokes equations). 
%\ti{Do we really need the acronyms PDE, ODE, NSE, FE, FD, FV, and SM?  I think we're only using them twice at most.  Since we already have a lot of acronyms, how about we get rid of these?}
Solving them allows %further 
prediction and analysis of the system's dynamics. The applicability of analytic methods for solving PDEs is usually limited to simple cases with special geometry and under severe assumptions. In practice, numerical approaches (e.g., finite difference, finite volume, spectral, and finite element methods) are utilized to discretize the governing equations and approximate the values of the unknowns %referring 
corresponding to a given grid. For turbulent %and non-homogeneous 
flows, we need to deal with an exceedingly large number of degrees of freedom due to the existence of a wide range of spatio-temporal scales to be resolved. Although such models, called here \emph{full order models} (FOMs), are capable of providing very accurate %and valuable 
results, they can be computationally demanding. Therefore, %they 
FOMs become impractical for applications %which 
that require multiple forward evaluations with varying inputs (e.g., flow control~\cite{noack2011reduced,ito1998reduced,ravindran2000reduced}, optimization~\cite{lang2009reduced,zahr2015progressive,degroote2010interpolation,amsallem2015design,benner2014model,bui2007goal,peherstorfer2018survey}, and digital twinning~\cite{rasheed2020digital,kapteyn2021probabilistic,hartmann2018model,tao2018digital,haag2018digital,boschert2016digital}) or studies requiring
several simulations like %in %\textit{Computational-Aided Clinical Trials}
computational-aided clinical trials 
\cite{VICECONTI2021120}.

\textit{Reduced order models (ROMs)} are defined as computationally light surrogates that can mimic the behavior of FOMs with sufficient accuracy~\cite{lucia2004reduced,rowley2017model,taira2017modal,taira2020modal,ahmed2021closures}. Projection-based ROMs have gained significant popularity in the past few decades due to the increased amounts of collected data (either from actual experiments or numerical simulations) as well as the development of system identification tools \cite{carlberg2017galerkin,swischuk2019projection}. Of particular interest, the combination of proper orthogonal decomposition (POD) and Galerkin projection has been a powerful driver for ROM %progresses. 
progress. The process comprises an \textit{offline} stage and an \textit{online} stage. The offline stage starts with the collection of data corresponding to %the system's 
system realizations (called \textit{snapshots}) at different time instants and/or parameter values. %This may involve a careful design of experiment process. 
With these data sets, POD provides a hierarchy of basis functions (or modes) that capture the maximum amount of the underlying system's energy (defined by the data variance). The offline stage is concluded by performing a Galerkin projection of the FOM operators onto the subspace spanned by a truncated set of POD modes to obtain a system of ordinary differential equations (ODEs) representing the Galerkin ROM (GROM). Although this offline stage can be extremely expensive, the resulting GROM can be utilized during the online deployment phase to efficiently predict the system's behavior at parameter values and/or time instants different from those in the data preparation process. 

The GROM framework has been successful in many applications (e.g., \cite{iollo2000stability,akhtar2009stability,sachs2010pod,noack2003hierarchy,bertagna2014model,yang2017efficient,hijazi2020data,girfoglio2021pod,ahmed2021closures}), especially those dominated by diffusion mechanisms or periodic dynamics. Those are often referred to as systems with a solution manifold that is characterized by a small Kolmogorov $n$-width \cite{ahmed2020breaking,peherstorfer2022breaking}. In the POD context, this means that the dynamics can be accurately represented by a few modes. However, for convection-dominated flows with strong nonlinearity, the Kolmogorov $n$-width is often large with a slow decay, which hinders the linear reducibility of the underlying system. 
%\AV{I am sorry, but I do not understand the next sentence, because 
%the GROM happens in the solved regime, isn't it? Then, this is not enough,
%but this is another story... do we need that sentence?}
%
%In other words, solution of GROM takes place in the \emph{under-resolved} regime. 

%The repercussions of a Galerkin truncation and projection is two-fold. 
%First, the representability of the resulting subspace (i.e., approximating the state as a linear superposition of the retained POD basis functions) is 
%\AV{Sorry, but the word weakened is also confusing for me: weak for me refers to the formulation of a problem or to some convergence.}
%reduced, giving rise to the projection error~\cite{singler2014new,amsallem2016pebl,ahmed2020long}. 
%\AV{I try to reformulate the previous sentence, as I understand it.}

The repercussions of a Galerkin truncation and projection %is 
are two-fold. First, the span of the retained POD basis functions does not necessarily provide an accurate representation of the solution and it gives rise to the 
\textit{projection error}~\cite{singler2014new,amsallem2016pebl,ahmed2020long}. Second, the interactions between the truncated and the retained modes can be significant. These interactions are ignored in the Galerkin projection step, and consequently the GROM cannot in general capture the dynamics of the resolved modes accurately. This introduces a \textit{closure error}~\cite{wang2012proper,sapsis2013blending,san2014proper,san2015stabilized,san2018extreme,xie2018data,pan2018data,rahman2019dynamic,imtiaz2020nonlinear,ahmed2020reduced,gupta2021neural}. Several efforts have been devoted to address the closure problem. %, please refer to our 
A recent survey covering a %span 
plethora of physics-based and data-driven ROM closure methodologies can be found in~\cite{ahmed2021closures}. 

The closure problem has been historically related to the stabilization of the ROM solution, drawing roots from large eddy simulation (LES) studies,  where the truncated small scales are thought of having diffusive effects on the larger scales. Therefore, eddy viscosity-based frameworks have been %prominent in ROM community. 
often used in the ROM literature~\cite{holmes2012turbulence}.
Nonetheless, it was found that introducing eddy viscosity to \emph{all} resolved scales can actually unnecessarily contaminate the dynamics of the \emph{largest} scales. To mitigate this problem, the \textit{variational multiscale (VMS)} method, 
which was proposed by Hughes' group~\cite{hughes1998variational,hughes2000large,hughes2001large} in the finite element setting (see, e.g.,~\cite{codina2018variational,john2016finite} for a survey), 
%~\cite{iliescu2013variational,iliescu2014variational} provides a clever way for addressing different levels of resolved and unresolved scales. For example, it 
was utilized to add eddy viscosity dissipation to only a portion of the ROM resolved scales in~\cite{iliescu2013variational,iliescu2014variational,wang2012proper}. A data-driven version of VMS (DD-VMS) has been recently proposed in~\cite{mou2021data}, where the effects of the truncated modes onto the GROM dynamics are not restricted to be diffusive.

%\subsection{Significance}
In the present study, we transform the DD-VMS~\cite{mou2021data,koc2021verifiability} and provide an alternative modular framework by utilizing machine learning (ML) capabilities. 
%\ti{I think we should be more assertive here.  We should say that this is a fundamental change in which the standard regression is replaced by ML.  And we should include some of this in the abstract. }
We stress that this is a fundamental change in which the standard DD-VMS regression is replaced by ML in order to better account for closure effects. Therefore, %we emphasize that 
the proposed neural network approach is %quite 
essentially different from the regression based %data-driven VMS-ROM approach 
DD-VMS \cite{mou2021data}. 
In particular, the DD-VMS ansatz of a quadratic polynomial closure model %- previously postulated - 
is relieved by utilizing the deep neural network (DNN) functionality with memory embedding. %We adopt 
We also leverage the long short-term memory (LSTM) variant of recurrent neural networks (RNNs) to approximate scale-aware closures. In essence, the use of LSTM encompasses a non-Markovian closure, supported by the Mori-Zwanzig %formulation
formalism~\cite{mori1965transport,zwanzig1980problems,chorin2000optimal,chorin2002optimal,chorin2009stochastic}. Moreover, we adopt the physics guided machine learning (PGML) framework introduced in~\cite{pawar2021physics,pawar2021model,pawar2021multi} to reduce the uncertainty of the output results. In particular, we exploit concatenation layers informed by the %VMS-GROM 
VMS-ROM arguments to enrich the neural network architecture and constrain the learning algorithm to the manifold of physically-consistent solutions. 
Finally, for problems with a large Kolmogorov $n$-width, 
we utilize the nonlinear POD (NLPOD) methodology~\cite{ahmed2021nonlinear} to reduce the projection error without affecting the computational efficiency,
by learning the correlations among the small unresolved scales to provide much fewer latent space variables. %We address 
%on a numerical test the different errors resulting from the Galerkin POD approach in our hybrid framework that benefits from the locality of scale interactions and information transfer, which is %one of the cornerstones of the VMS method.
%For the two-dimensional vortex merger problem, we 
We also perform a numerical investigation of the proposed strategies (ML-VMS-ROM, PGML-VMS-ROM, and NLPOD-VMS-ROM), with a particular focus on the locality of scale interactions, which is a cornerstone of the VMS framework.

%\subsection{Structure of the paper}

The rest of the paper is organized as follows: %. 
We briefly describe the reduced order modeling methodology by the nexus of POD and Galerkin projection in~\cref{sec:rom}. The relevant background information and notations %from 
for the VMS approach are given in~\cref{sec:vms}. The use of the PGML methodology to provide reliable predictions is explained in~\cref{sec:pgml}, while the NLPOD approach is discussed in~\cref{sec:nlpod}. The proposed NLPOD-PGML-VMS framework is tested for the parametric unsteady vortex-merger problem, which %exemplify 
exemplifies convection-dominated flow systems. Results and discussions are presented in~\cref{sec:res}, followed by the concluding remarks in~\cref{sec:conc}.

\section{Reduced Order Modeling} \label{sec:rom}
%%===============================================%%
% \AV{I have nothing against the use of NSE as model problem. However,
% we do numerical experiment on the Vorticity-Stream function formulation.
% So, to be consistent, I see two options: (i) we introduce vorticity-sf here,
% (ii) we add a numerical test on NSE.}
% \ti{I like (i).}

A Newtonian incompressible fluid flow in a domain $\Omega \subset \mathbb{R}^d$, where $d$ defines the spatial dimension (i.e., $d\in\{2,3\})$, can be described by the Navier-Stokes equations (NSE). In order to %mitigate the odd-even decoupling problem and 
eliminate the pressure term, we consider the NSE in the vorticity-vector potential formulation. In particular, we consider the 2D case where the vector potential is reduced to the streamfunction as follows:
\begin{equation}
\begin{aligned}
\partial_t\omega - \nu \Delta \omega + (\bu \cdot \nabla)\omega &= 0, \qquad \text{in} \ \Omega \times [0,T], \\
      \Delta \psi + \omega &= 0, \qquad \text{in} \ \Omega \times [0,T],
\end{aligned} \label{eq:NSE}
\end{equation}
where $\omega(\bx,t)$ and $\psi(\bx,t)$ denote the vorticity and streamfunction fields, respectively, for $\bx \in \Omega$ and $t \in [0, T]$, while $\nu$ stands for the kinematic viscosity (diffusion coefficient). In dimensionless form, %this 
$\nu$ represents the reciprocal of the Reynolds number, $\re$. The velocity vector field $\bu(\bx,t)$ is related to the streamfunction as follows:
\begin{equation}
    \bu = \nabla ^{\perp} \psi, \quad \nabla ^{\perp} = [\partial_y, -\partial_x]^T.
    \label{eqn:u-psi}
\end{equation}
%\ti{Omer, should it be $\bu = - \nabla ^{\perp}  \psi$?  This way the Jacobian in the next equation is consistent with the definition.  Could you please check?}

By using Eq.~\eqref{eqn:u-psi}, %Equation
Eq.~\cref{eq:NSE} can be further rewritten %by eliminating the vorticity field vector 
as follows:
\begin{equation}
\begin{aligned}
\partial_t\omega - \nu \Delta \omega + J(\omega,\psi) &= 0, \qquad \text{in} \ \Omega \times [0,T],
\end{aligned} \label{eq:NSEv}
\end{equation}
where $J(\cdot,\cdot)$ denotes the Jacobian operator, which is defined as follows:
\begin{equation}
    J(\omega,\psi) = \dfrac{\partial \omega}{\partial x} \dfrac{\partial \psi}{\partial y} -  \dfrac{\partial \omega}{\partial y} \dfrac{\partial \psi}{\partial x}.
\end{equation}

The vorticity transport equation (Eq.~\cref{eq:NSEv}) is equipped with an initial condition %at $t=0$ 
and boundary conditions on $\Gamma:= %\delta 
\partial \Omega$. For convenience and simplicity of presentation, we shall assume the following conditions:
\begin{equation}
\begin{aligned}
IC:& \ \omega(\bx, 0) = \omega_0(\bx), & \text{in} \ \Omega, \\
BC \ \mathrm{(non-slip)}:& \ \psi(\bx,t) = 0, \quad \dfrac{\partial \psi}{\partial \boldsymbol{n}} = 0, & \text{in} \ \Gamma \times [0,T].
\end{aligned} \label{eq:ic_bc}
\end{equation}
In the %remaining 
remainder of this section, we describe the %process of obtaining a 
construction of the projection-based ROM of the vorticity transport equation. This includes the use of POD to %express an 
approximate the solution %subspace in 
(\cref{sec:pod}), followed by the Galerkin method, where the FOM operators in Eq.~\cref{eq:NSE} are projected onto the POD subspace to define the sought GROM (\cref{sec:gp}).

%%-----------------------------------------------------------%%
\subsection{Proper orthogonal decomposition} \label{sec:pod}
%%-----------------------------------------------------------%%
We consider a collection of %the system's 
system realizations defined by an ensemble of %the 
vorticity fields %data $\mathcal{W}:= \text{span}
$\{\omega(\bx,t_0),\allowbreak \omega(\bx,t_1),\dots, \allowbreak \omega(\bx,t_{M-1})\}$. These are often called \emph{snapshots} and come from either experimental measurements or numerical simulations of Eq.~\cref{eq:NSE} or Eq.~\cref{eq:NSEv} using any of the standard discretization schemes (e.g., finite element, finite difference or finite volume methods).
Without loss of %generalization, 
generality, we assume that these snapshots %data 
are sampled at equidistant $M$ $(>1)$ time instants with $t_m = m\Delta t$, where $m=0,1,\dots, M-1$ and $\Delta t = \dfrac{T}{M-1}$. We note that, in general, these snapshots can correspond to different types of parameters %as well
(e.g., operating conditions, physical properties, and geometry).

In POD, we seek a low-dimensional basis $\{\phi_1,\phi_2,\dots,\phi_R\}$ %in $\mathcal{H}$ 
that optimally approximates %$\mathcal{W}$ 
the space spanned by the snapshots
in the following sense~\cite{holmes2012turbulence}:
%\AV{If time is the parameter we consider, it is ok to emphasize the dependence on $t_m$. However, here we would like to have a more general breath 
%to my understanding, so the average could be taken with respect to another parameter - viscosity? geometry? If this is the case, I would drop
%the $t_m$ hereafter.}
\begin{equation}
\begin{aligned}
    &\min \ \Bigg\langle \bigg\| \omega(\cdot,\cdot) - \sum_{k=1}^{R} \big(\omega(\cdot,\cdot),\phi_k(\cdot)\big) \phi_k(\cdot) \bigg\|^2 \Bigg\rangle, \\
    &\text{subject to}  \qquad \quad \| \phi \| = 1,  \qquad \big(\phi_i(\cdot),\phi_j(\cdot)\big) = \delta_{ij} ,
\end{aligned} \label{eq:opt}
\end{equation}
where $\langle \cdot \rangle$ denotes an average operation with respect to the parametrization, %. 
$(\cdot,\cdot)$ is an inner product, and $\| \cdot \|$ is the corresponding norm.
For example, an ensemble average based on temporal snapshots %reads:
can be defined as follows:
\begin{equation}
    \langle \omega \rangle = \dfrac{1}{M} \sum_{m=0}^{M-1} \omega(\cdot,t_m).
\end{equation}

The snapshots represent the approximation of the quantity of interest on a specific grid. For example, a realization of the vorticity field at a given time can be arranged in a column vector $\boldsymbol{\omega} \in \mathbb{R}^N$, where $N$ is the number of grid points. %Following that, it 
It can be shown that solving the optimization problem~\cref{eq:opt} amounts to solving the following eigenvalue problem~\cite{volkwein2013proper}:
\begin{equation}
    \mathbf{D} \Phi =  \Phi \Lambda, \label{eq:eig_sp}
\end{equation}
where the entries of the diagonal matrix $\Lambda$ and the columns of $\Phi$ represent the eigenpairs of the spatial autocorrelation matrix $\mathbf{D} \in \mathbb{R}^{N\times N}$ with entries defined as
\begin{equation}
    \big[\mathbf{D}\big]_{ij} = \bigg\langle \boldsymbol{\omega}(\bx_i,\cdot) \boldsymbol{\omega}(\bx_j,\cdot) \bigg\rangle,
\end{equation}
where $\boldsymbol{\omega}(\bx_i,\cdot)$ is the $i$-th entry of $\boldsymbol{\omega}$. For fluid flow problems, the length of the vector $\boldsymbol{\omega}$ is %significantly 
often large, %(i.e., $N \gg 1$), 
which makes the eigenvalue problem in Eq.~\cref{eq:eig_sp} computationally %intractable.
challenging.

Sirovich~\cite{sirovich1987turbulence1,sirovich1987turbulence2,sirovich1987turbulence3} proposed a numerical procedure, known as the \emph{method of snapshots}, to reduce the computational cost of solving Eq.~\cref{eq:eig_sp}. This approach is efficient, especially when the number of collected snapshots $M$ is much smaller than the number of degrees of freedom (i.e., $M\ll N$), as it reduces the $N\times N$ eigenvalue problem in Eq.~\cref{eq:eig_sp} to an $M\times M$ problem. The spatial autocorrelation matrix $\mathbf{D} \in \mathbb{R}^{N\times N}$ is replaced by the temporal snapshot correlation matrix $\mathbf{K} \in \mathbb{R}^{M\times M}$ with entries defined as follows:
%\AV{Following up my previous check, is the autocorrelation defined
%so to subtract the average of the snapshots? I think so, but we should probably %mention this. By the way, again here we write the example with time as the parameter, %but we do not stress that we mau have more parameters.}
\begin{equation}
    \big[\mathbf{K}\big]_{ij} = \dfrac{1}{M} \bigg(\omega(\cdot,t_i), \omega(\cdot,t_j) \bigg) .
\end{equation}
The following eigenvalue problem is thus considered:
\begin{equation}
    \mathbf{K} \mathbf{v}_k =  \lambda_k \mathbf{v}_k, \label{eq:eig_tmp}
\end{equation}
where $\mathbf{v}_k$ is the $k^{\text{th}}$ eigenvector of $\mathbf{K}$ and $\lambda_k$ is the associated eigenvalue. To obtain the hierarchy of the POD basis, the eigenpairs are sorted in a descending order by their eigenvalues (i.e., $\lambda_1 \ge \lambda_2 \dots \ge \lambda_M \ge 0$). Finally, the POD basis functions can be computed as a linear superposition of the collected snapshots as follows~\cite{volkwein2013proper}:
\begin{equation}
    \phi_k(\cdot) = \dfrac{1}{\sqrt{\lambda_k}} \sum_{m=0}^{M-1} [\mathbf{v}_k]_m \omega(\cdot,t_m), \label{eq:pod_basis}
\end{equation}
where $[\mathbf{v}_k]_m$ denotes the $m^{\text{th}}$ component of $\mathbf{v}_k$. It can be verified that the basis functions in Eq.~\cref{eq:pod_basis} are orthonormal (i.e., $(\phi_i(\cdot),\phi_j(\cdot))=\delta_{ij}$), where $\delta_{ij}$ is the Kronecker delta.
%, defined as follows:
%\begin{equation}
%    \delta_{ij} = 
%    \begin{cases}
%     1, \quad \text{if} \ i=j,\\
%     0, \quad \text{otherwise}.
%    \end{cases}
%\end{equation}
The POD eigenvalues define the contribution of each mode toward the total variance in the given snapshots.
%For the NSE where the state variables are the velocity components, this corresponds to the amount of (kinetic) energy captured by each mode. 
%\ti{For the vorticity-streamfunction formulation this should be rephrased.}
A metric that evaluates the quality of a given set of retained modes in representing the system is the relative information content (RIC)~\cite{ahmed2021closures}, defined as follows:
\begin{equation}
    \text{RIC}(k) = \dfrac{\sum_{l=1}^{k}\lambda_l}{\sum_{l=1}^{M}\lambda_l}, \label{eq:ric}
\end{equation}
where $k$ is the POD index at which modal truncation takes place. We emphasize that the same approach can be applied considering %different 
parameters other than time. In this case, the temporal correlation matrix %may be 
is substituted by a generalized parameter correlation matrix.

\subsection{Galerkin projection} \label{sec:gp}
%%--------------------------------- ------------%%
% The weak (variational) formulation of the vorticity-transport equation can be written as follows\cite{liu2001simple}: find $\omega \in H^{1}(\Omega)$ and $\psi \in H^{1}_0(\Omega)$ such that $\omega(\bx,0)=\omega_0(\bx)$ and
% \begin{equation} \begin{aligned}
% (\partial_t\omega,\zeta) - \nu (\Delta \omega, \zeta)  + \big(J(\omega,\psi), \zeta \big) &= 0, \qquad \forall \zeta \in \bX.%\\ 
% %(\Delta \psi, \xi) &= -(\omega, \xi), \ \quad \forall \xi \in Q.
% \end{aligned} \label{eq:NSE2}\end{equation}
%\ti{Omer, I've commented out the first paragraph of this section, including the equation with the weak formulation.  The reason is that we are keeping the \textbf{Laplacian} in the weak formulation.  This is different from what Liu and E do in [38] since they use the divergence theorem and get $(\nabla w , \nabla \zeta)$. (BTW, please add E to the list of authors in [38].) To use the Laplacian, we need more regularity than $H^1$.  Thus, in this section I tried to get rid of as much mathematical notation/spaces/formulas/etc as possible and keep only the material that we actually need for the GROM.  Could you please check?}

The GROM starts by the Galerkin truncation step, making use of the optimality criterion in Eq.~\cref{eq:opt} as follows:
\begin{equation}
    \omega(\bx,t_m) \approx \omega_R(\bx,t_m) = \sum_{k=1}^R a_k(t_m) \phi_k(\bx), \label{eq:upod}
\end{equation}
where $\{a_k\}_{k=1}^{R}$ are the time-varying modal coefficients (weights), known as \textit{generalized coordinates}. The optimal values of these coefficients are defined by the true projection of the FOM trajectory onto the corresponding POD basis function as follows:
\begin{equation}
    a_k(t_m) = \big(\omega(\cdot,t_m),\phi_k(\cdot)\big). \label{eq:truepod}
\end{equation}
%The 
Next, the vorticity field $\omega$ in %Eq.~\cref{eq:NSE2} 
Eq.~\cref{eq:NSEv} is replaced by its approximation $\omega_R$ from Eq.~\cref{eq:upod}. After this, the Galerkin projection step comes into play, by defining the POD test subspace $\bX_R$ %\subseteq \bX$ 
as follows:
\begin{equation}
    \bX_R := \text{span}\{\phi_1, \phi_2, \dots, \phi_R \}.
\end{equation}
Finally, Eq.~\cref{eq:NSEv} with $\omega$ replaced  by $\omega_R$ is projected onto the POD space $\bX_R$.
This yields the GROM of the vorticity transport equation: %reads as follows: 
Find $\omega_R \in \bX_R$ such that:
\begin{equation}\begin{aligned}
(\partial_t\omega,\phi) - \nu (\Delta \omega, \phi)  + \big(J(\omega,\psi), \phi \big) &= 0,  \qquad \forall \phi \in \bX_R.\\ 
\end{aligned} \label{eq:GNSE} \end{equation}
Equation~\cref{eq:GNSE} can be written in a tensorial form as follows:
\begin{equation}
    \dot{\ba} = A\ba + \ba^\top B \ba, \label{eq:Gtens}
\end{equation}
where $\ba(t) \in \mathbb{R}^R$ is the vector of unknown coefficients $\{a_k\}_{k=1}^{R}$, while $A \in \mathbb{R}^{R\times R}$ %matrix 
and $B \in \mathbb{R}^{R\times R\times R}$ are the matrix and tensor corresponding to the linear and nonlinear terms, respectively. 

%We note that the pressure term vanishes in Eq.~\cref{eq:GNSE} thanks to the assumption that the POD basis functions are discretely divergence-free (solenoidal) and satisfy the boundary conditions given in Eq.~\cref{eq:ic_bc} as follows:
% \begin{equation} 
%     \int_{\Omega} \nabla p \bphi \ \dd \bx = \underbrace{\cancel{\int_{\Gamma} p \bphi\cdot \mathbf{n} \ \dd \gamma}}_{\text{boundary conditions}} - \underbrace{\cancel{\int_{\Omega} p \nabla \cdot \bphi \ \dd \bx}}_{\text{divergence free}} = 0.
% \end{equation}

%In order to achieve computational gains, the ROM usually operates in the under-resolved regimes (i.e., $R \ll N$). First, 
The Galerkin truncation step restricts the approximation of the %velocity 
vorticity field to live in a low-rank subspace $\bX_R$ ($R \ll N$), which might not capture all the relevant flow structures. Therefore, a projection error is introduced. %Then, 
Furthermore, the Galerkin projection step enforces the dynamics of the ROM to be defined using only the scales supported by $\bX_R$. Nonetheless, due to the coupling between different modes, the unresolved scales (i.e., the scales modeled by $\{\phi_k\}_{k\geq R+1}$) influence the dynamics of the resolved scales (i.e., the scales modeled by $\{\phi_k\}_{k\leq R}$). By neglecting these mutual interactions, the GROM becomes incapable of accurately describing the dynamics of the retained modes, which is usually referred to as the \textit{closure} problem~\cite{ahmed2021closures}.

\begin{figure}[h]
\begin{center}
	\includegraphics[width=0.8\textwidth]{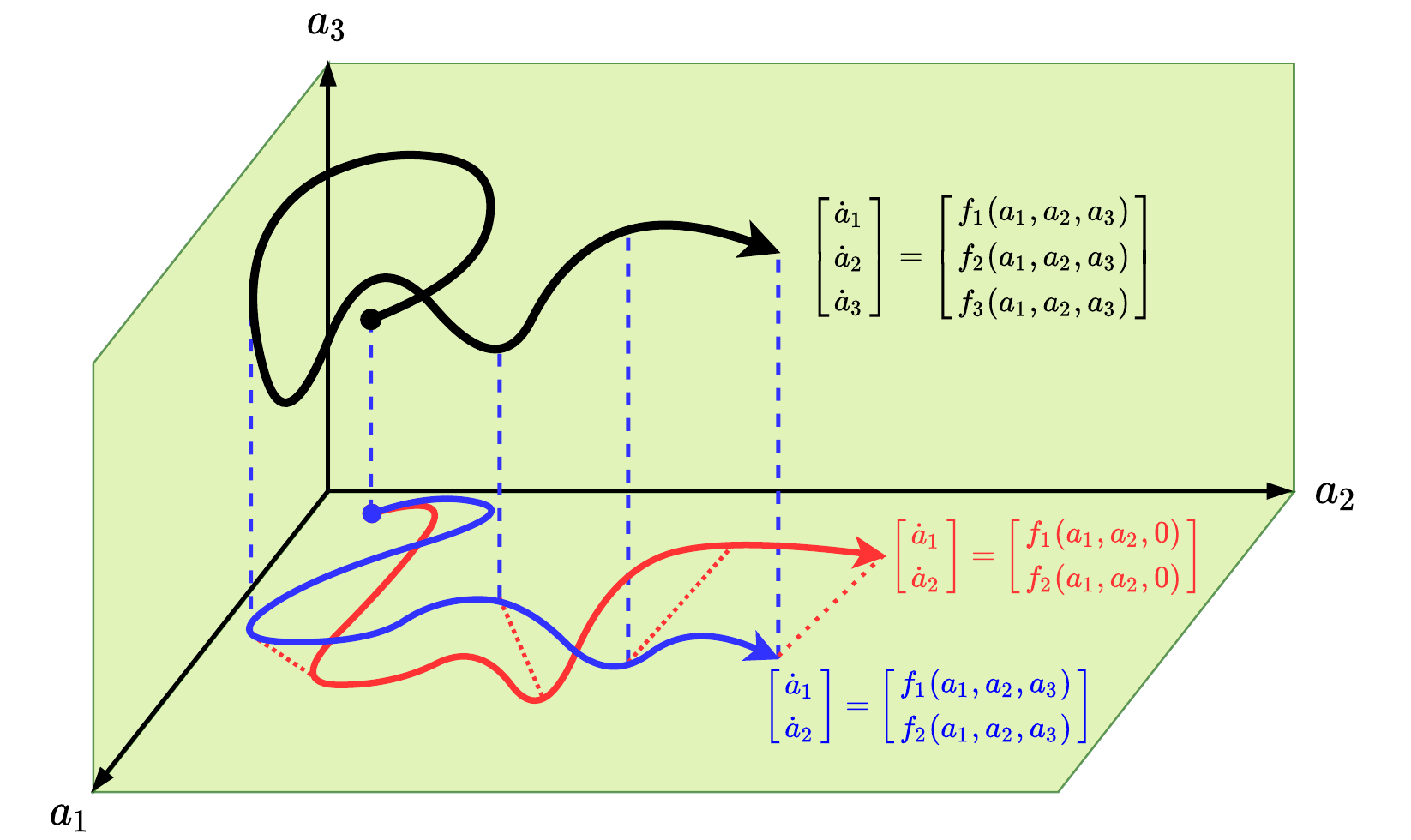}
    \caption{Representation of the repercussions of modal truncation onto the ROM solution. The solid black curve denotes the FOM trajectory, assuming that the full rank expansion is defined by $a_1$, $a_2$, and $a_3$. The solid blue curve defines the projection of the FOM trajectory onto a two-dimensional subspace. The vertical dashed blue lines refer to the \textrm{projection or representation error}. Note that evaluating $a_1$ and $a_2$ still requires the knowledge of the FOM trajectory (i.e.,$a_1$, $a_2$, and $a_3$) at every point. In practice, we only have information regarding the resolved variables (i.e., $a_1$ and $a_2$), so the contribution of $a_3$ towards the dynamics of $a_1$ and $a_2$ is neglected. This yields a \textrm{closure error}, denoted by the dashed red lines.} %in this figure.}
	\label{fig:error}
\end{center}
\end{figure}

The projection error and closure error are illustrated in~\cref{fig:error}, for a toy system whose full-rank approximation can be %expanded 
represented with 3 modes as follows: 
\begin{equation}
    \omega(x,t) = a_1(t) \phi_1(x) + a_2(t) \phi_2(x) + a_3(t) \phi_3(x). \label{eq:toy1}
\end{equation}
Assuming that the FOM is written in the following form:
\begin{equation}
    \dot{\omega} = F(\omega), \label{eq:udottoy}
\end{equation}
then the dynamics of $\{a_k\}_{k=1}^{3}$ can be described as $\dot{a}_k = (F(\omega),\phi_k)$. Thus, the FOM trajectory can be written as follows:
\begin{equation}
\begin{aligned}
    \begin{bmatrix} \dot{a}_1\\ \dot{a}_2\\ \dot{a}_3 \end{bmatrix} = \begin{bmatrix}     f_1(a_1,a_2,a_3)\\f_2(a_1,a_2,a_3)\\f_3(a_1,a_2,a_3) \end{bmatrix}.
\end{aligned} \label{eq:toy2} 
\end{equation}
In other words, evolving $\{a_k\}_{k=1}^{3}$ using Eq.~\cref{eq:toy2} and reconstructing %$u$ 
$\omega$ with Eq.~\cref{eq:toy1} recovers the FOM field (equivalent to solving Eq.~\cref{eq:udottoy} using standard discretization schemes). For the sake of demonstration, we suppose that we retain only 2 modes in the ROM approximation. This corresponds to removing the third row in Eq.~\cref{eq:toy2} as follows:
\begin{equation}
\begin{aligned}
    \begin{bmatrix} \dot{a}_1\\ \dot{a}_2 \end{bmatrix} = \begin{bmatrix} f_1(a_1,a_2,a_3)\\f_2(a_1,a_2,a_3) \end{bmatrix}. 
\end{aligned} \label{eq:toy3} 
\end{equation}
Approximating %$u$ 
$\omega$ with just two modes results in losing the flow structures that are contained in the truncated mode (the vertical direction in~\cref{fig:error}), which %defines 
yields the projection error. %Nonetheless, we note 
Furthermore, we note 
that %$f_k$ is usually a function 
$f_1$ and $f_2$ are usually functions 
of $a_1$, $a_2$, and $a_3$ for systems with strong nonlinearity and coupling between different modes. However, during ROM deployment, we do not usually have information regarding the unresolved dynamics ($a_3$ in this example). %In 
Thus, in GROM, the effects of the truncated scales onto the resolved scales are assumed to be negligible, as follows:
\begin{equation}
\begin{aligned}
    \begin{bmatrix} \dot{a}_1\\ \dot{a}_2 \end{bmatrix} = \begin{bmatrix} f_1(a_1,a_2,0)\\f_2(a_1,a_2,0) \end{bmatrix}.
\end{aligned} \label{eq:toy4} 
\end{equation}
We denote the reference trajectory described by Eq.~\cref{eq:toy3} as the true projection, which is %equivalent 
related to Eq.~\cref{eq:truepod}. This defines the best low-rank approximation that can be obtained for a given number of modes, assuming we have access to the whole set of FOM scales. The difference between the GROM trajectory (corresponding to solving Eq.~\cref{eq:toy4}) and the true projection trajectory represents the closure error. %We address both types of errors in the present study. 
In the present study, we address both the closure error and the projection error.
%First, we consider the closure problem by developing the PGML methodology in \cref{sec:pgml} to model the effects of truncated modes onto the ROM dynamics, supported by the VMS framework in \cref{sec:vms}. 
First, to tackle the closure problem, we leverage the VMS framework outlined in \cref{sec:vms} to develop the PGML methodology in \cref{sec:pgml}.
Then, we utilize the NLPOD approach in \cref{sec:nlpod} to reduce the projection error by learning a compressed latent space that encapsulates some of the truncated flow structures.

\section{Variational Multiscale Method} \label{sec:vms}
%%======================================================%%
%\,

%\ti{I think we need to change a bit the presentation in  this section.}
The variational multiscale (VMS) methods are general numerical discretizations that significantly increase the accuracy of classical Galerkin approximations in under-resolved simulations, e.g., on coarse meshes or when not enough basis functions are available. The VMS framework, which was proposed by Hughes and coworkers~\cite{hughes1998variational,hughes2000large,hughes2001large}, has made a profound impact in many areas of computational %mathematics 
mechanics (see, e.g.,~\cite{codina2018variational,john2016finite} for a survey).

To illustrate the standard VMS methodology, we consider a general nonlinear %system/PDE:
partial differential equation 
\begin{equation}
    \dot{\omega} = \bFF(\omega), \label{eq:udot}
\end{equation}
whose weak (variational) form is %:
\begin{equation}
    (\dot{\omega},\phi) = (\bFF(\omega),\phi), \quad \forall \phi \in \bX,
\end{equation}
where $\bFF$ is a general nonlinear function and $\bX$ is an appropriate test space. To build the VMS framework, we start with a sequence of hierarchical spaces of increasing resolutions: $\bX_1$, $\bX_1 \oplus \bX_2$, $\bX_1 \oplus \bX_2 \oplus \bX_3$, $\dots$. Next, we project system~\cref{eq:udot} onto each of the spaces $\bX_1$, $\bX_2$, $\bX_3$, $\dots$, which yields a separate equation for each space. %The 
From a computational efficiency point of view, the goal is to solve for the $\omega$ component that lives in the coarsest space (i.e., $\bX_1$), since this yields the lowest-dimensional system:
\begin{equation}
    (\dot{\omega},\phi) = (\bFF(\omega),\phi), \quad \forall \phi \in \bX_1. \label{eq:ux1}
\end{equation}
However, system~\cref{eq:ux1} is \emph{not} closed since its right-hand side involves $\omega$ components that do not %live in 
belong to $\bX_1$ (i.e., $\omega_2 \in \bX_2$, $\omega_3 \in \bX_3$, $\dots$):  %as follows:
%\AV{Why in the following equation $\bu_1$ and $\bu_2$ are summed and $\bu_3$ is not?}
\begin{equation}
   (\bFF(\omega),\phi) = (\bFF(\omega_1,\omega_2,\omega_3,\dots),\phi), \quad \forall \phi \in \bX_1. \label{eq:rhsx1}
\end{equation}

% \AV{The following sentence to me is a little bit vague in that ``mainly''.
% If the solution is developed on an orthogonal basis and the problem is linear,
% then the closure problem is trivial. If the problem is linear, but the basis is not orthogonal, we still have a closure to address, isn't it? I agree that, overall, the nonlinearity is the main contributor, but I think we should be more accurate.}
% \ti{This is an excellent point.  Even if the problem is linear, we may still have closure.  We've shown this in~\cite{koc2019commutation}, where we called it the ``commutation error."  But in all fairness, closure is spectacular for nonlinear problems and relatively benign for linear problems.  So most people don't care about closure for linear problems. 

% In any case, I agree with Ale that we should be more accurate.  How about we just eliminate the next sentence?}
% This coupling is mainly due to the nonlinearity of $\bFF$. 

Thus, the VMS closure problem needs to be solved. That is, Eq.~\cref{eq:rhsx1} needs to be %approximated in $\bX_1$.
replaced with an equation that involves only terms that belong to $\bX_1$.
In general, the VMS system in Eq.~\cref{eq:ux1} equipped with an appropriate closure model (i.e., a model with components in $\bX_1$ that captures the interaction between $\omega_1$ and the scales in $\bX_2, \bX_3, \ldots$) yields an accurate approximation of the $\bX_1$ component of $\omega$.

The POD procedure in~\cref{sec:pod} yields a hierarchy of orthogonal basis functions, sorted by their contribution to the total energy. Therefore, it provides a natural fit %into 
to the VMS framework. %We 
Next, we illustrate the adoption of VMS in GROM settings to define a multi-level VMS ROM. In particular, we detail the two-scale and the three-scale VMS ROMs, while further extensions become straightforward.

%%----------------------------------------------%%
\subsection{Two-scale VMS ROM} \label{sec:vms2}
%%----------------------------------------------%%
The two-scale VMS (VMS-2) ROM utilizes two orthogonal spaces, $\bX_1$ and $\bX_2$, defined as follows:
\begin{equation}
\begin{aligned}
    \bX_1 &:= \text{span}\{\phi_1, \phi_2, \dots, \phi_R \}, \\
    \bX_2 &:= \text{span}\{\phi_{R+1}, \phi_{R+2}, \dots, \phi_N \},
\end{aligned}
\end{equation}
where $\bX_1$ represents the span of the resolved ROM scales and $\bX_2$ is the span of the unresolved scales. Thus, $\omega$ can be written as follows:
\begin{equation}
    \omega = \sum_{k=1}^R a_k \phi_k + \sum_{k=R+1}^N a_k \phi_k = \underbrace{\omega_R}_{\text{resolved}} + \underbrace{\omega'}_{\text{unresolved}}, \label{eq:udecom2}
\end{equation}
where $\omega_R \in \bX_1$ is the resolved ROM component of $\omega$, while $\omega' \in \bX_2$ is the unresolved component. %Equation
Using this decomposition, Eq.~\cref{eq:ux1} can be rewritten as follows:
\begin{equation}
    \big(\dot{\omega}_R,\phi_k\big) = \big(\bFF(\omega_R),\phi_k\big) + \underbrace{%\boxed{
    \bigg[\big(\bFF(\omega),\phi_k\big) - \big(\bFF(\omega_R),\phi_k\big) \bigg]%}
    }_{\text{VMS-2 closure term}}, \  \forall k \in \{1,\dots,R\}. \label{eq:vms2}
\end{equation}
The %boxed 
bracketed term in Eq.~\cref{eq:vms2} is the VMS-2 closure term, which models the interaction between the ROM modes and the discarded modes. Since the unresolved component of $\omega$, %outside $\bX_1$
$\omega'$,
is not available during 
%\AV{is ``deployment'' a standard terminology for denoting the online stage?}
online deployment stage, it is not possible to exactly compute the closure term in practical settings. Instead, the closure term can be approximated using a generic function $\bGG(\omega_R)$ as follows:
\begin{equation}
    \big( \bGG(\omega_R), \phi_k) \approx \big(\bFF(\omega),\phi_k\big) - \big(\bFF(\omega_R),\phi_k\big),
\end{equation}
and the VMS-2 ROM can be written as
\begin{equation}
    \big(\dot{\omega}_R,\phi_k\big) = \big(\bFF(\omega_R),\phi_k\big) + \big( \bGG(\omega_R), \phi_k). \label{eq:vms22}
\end{equation}
The form and parameters of $\bGG$ will be defined in~\cref{sec:pgml}.

%%------------------------------------------------%%
\subsection{Three-scale VMS ROM} \label{sec:vms3}
%%------------------------------------------------%%
The \textit{locality of modal interactions} %and information transfer} 
is a cornerstone of the VMS framework. It states that neighboring modes have more mutual interactions than those who are far apart in the energy spectrum. For this reason, it %makes sense to have a fine distinction on the closure models 
is natural to distinguish between neighboring and far modes when closure modeling is performed. To this end, the %The 
flexibility of the hierarchical structure of the ROM space %allows, for instance, 
is leveraged to perform a three-scale decomposition of $\omega$, leading to a three-scale VMS (VMS-3) ROM, %that can be even more accurate than the VMS-2 ROM. 
which aims at increasing the VMS-2 ROM accuracy.
%We 
To construct the VMS-3 ROM, we
first build three orthogonal spaces, $\bX_1$, $\bX_2$, and $\bX_3$, as follows:
\begin{equation}
\begin{aligned}
    \bX_1 &:= \text{span}\{\phi_1, \phi_2, \dots, \phi_r \}, \\
    \bX_2 &:= \text{span}\{\phi_{r+1}, \phi_{r+2}, \dots, \phi_R \},\\
    \bX_3 &:= \text{span}\{\phi_{R+1}, \phi_{R+2}, \dots, \phi_N \}.
\end{aligned}
\end{equation}
Compared to the %distinction between the 
decomposition into resolved and unresolved scales in~\cref{sec:vms2}, $\bX_1$ now represents the \emph{large resolved} ROM scales, %and 
$\bX_2$ represents the \emph{small resolved} ROM scales, while $\bX_3$ denotes the unresolved ROM scales. With these definitions, $\omega$ can be written as follows:
\begin{equation}
\begin{aligned}
\omega &= \sum_{k=1}^r a_k \phi_k + \sum_{k=r+1}^R a_k \phi_k + \sum_{k=R+1}^N a_k \phi_k \\
        &= \underbrace{\omega_L}_{\text{large resolved}} + \underbrace{\omega_S}_{\text{small resolved}} + \underbrace{\omega'}_{\text{unresolved}}.
    \end{aligned}
\end{equation}
This is similar to Eq.~\cref{eq:udecom2} with $\omega_R = \omega_L + \omega_S$. To construct the VMS-3 ROM, %we repeat Eq.~\cref{eq:ux1} for $\bX_1$ and $\bX_2$ as follows:
we project system~\cref{eq:udot} onto each of the spaces $\bX_1$ and $\bX_2$, as follows:
\begin{align}
    \big(\dot{\omega}_L,\phi_k\big) = \big(\bFF(\omega_L+\omega_S),\phi_k\big) + %\boxed{
    \bigg[\big(\bFF(\omega),\phi_k\big) - \big(\bFF(\omega_L+\omega_S),\phi_k\big) \bigg], %}, 
    \  k = 1,\ldots,r, \label{eq:vms31} \\
    \big(\dot{\omega}_S,\phi_k\big) = \big(\bFF(\omega_L+\omega_S),\phi_k\big) + %\boxed{
    \bigg[\big(\bFF(\omega),\phi_k\big) - \big(\bFF(\omega_L+\omega_S),\phi_k\big) \bigg], %}, 
    \  k = r+1,\ldots,R. \label{eq:vms32} 
\end{align} 
Although the two %boxed 
bracketed terms in Eq.~\cref{eq:vms31} and Eq.~\cref{eq:vms32} defining the VMS-3 closure terms look similar, they have different roles. The first term models basically 
%\av{I SEE THAT "BASICALLY" WAS COMMENTED, OK, HOWEVER, STRICTLY SPEAKING THIS TERM MODELS THE ACTION OF THE SMALL AND UNRESOLVED TO THE LARGE - THEN BY THE LOCALITY PRINCIPLE WE ARGUE THAT THE INTERACTION OF THE LARGE AND THE UNRESOLVED IS MINIMAL, IS THIS CORRECT? THE "BASICALLY" WAS INTENDED TO SUMMARIZE THIS. ANYWAY, IF YOU DON'T LIKE IT, NO PROBLEM. }
%\ti{And now "basically" is baaaack :)}
the interaction between the large  and the small resolved modes, because the interaction large-unresolved is assumed to be negligible (according to the VMS principle of locality of modal interactions). The second term models %, in fact, 
the interaction between the small resolved  and the unresolved ROM modes. This allows great flexibility in choosing the structure of the different VMS ROM closure terms. 
%This is in agreement with the locality of energy transfer, which is one of the main concepts of the VMS framework. This implies that the neighboring modes having more interactions than those who are far apart in the energy spectrum. Therefore, different levels might require different closure models. 
This concept is %validated 
investigated numerically in~\cref{sec:res}.

%%=========================================================%%
\section{Physics Guided Machine Learning} \label{sec:pgml}
%%=========================================================%%
In this section, the %The 
VMS-2 and VMS-3 closure terms defined in~\cref{sec:vms} are approximated using only the information in the resolved scales. Specifically, we %We 
utilize a purely data-driven approach to compute the parameters of the closure models. %Instead 
However, instead of relying on heuristics or ad-hoc arguments to define the specific structure of the closure model (as in the standard DD-VMS~\cite{mou2021data}), we exploit the capabilities of deep neural network (DNN) in approximating arbitrary functions. %can be exploited to do this task. 
In particular, we use the long short-term memory (LSTM) variant of recurrent neural networks (RNNs), which has shown substantial success in data-driven modeling of time series \cite{gers2002applying,siami2019performance,hua2019deep}. %\AV{Do we have references for the statement above?}
%\ti{I agree.  We should add some refs.}
We emphasize that, to mitigate well-known drawbacks of data-driven modeling (e.g., sensitivity to noise in input data), the VMS ROM framework utilizes data to model only the VMS ROM closure operators, %. All 
but all 
the other ROM operators are built by using classical Galerkin projection. Thus, our VMS ROM framework incorporates ``data-driven closure,'' %,
rather than ``data-driven modeling'' for the resolved scales.

\subsection{ML-VMS ROM}
The VMS-2 ROM in Eq.~\cref{eq:vms22} can be rewritten as follows:
\begin{equation}
    \dot{\ba} = \bff(\ba) + \bc(\ba),
    \label{eqn:ml-vms-2-1}
\end{equation}
where $\ba = [a_1, a_2, \dots, a_R]^T \in \mathbb{R}^R$ is the vector of coefficients for resolved POD modes, $\bff(\ba) = [\big(\bFF(\omega_R),\phi_1\big), \big(\bFF(\omega_R),\phi_2\big), \dots, \big(\bFF(\omega_R),\phi_R\big)]$ represents the Galerkin projection of the FOM operators onto the POD subspace, and $\bc(\ba) = [c_1, c_2, \dots, c_R]^R\in \mathbb{R}^R$ is the vector of the closure (correction) terms, i.e., $c_k = (\bGG(\omega_R),\phi_k)$. In the present study, we use DNN to represent the closure model, i.e.,  $\bc(\cdot) \approx \pi_{\theta}(\ba)$, where $\theta$ denotes the parameterization of the LSTM. The general functional form of the DNN models used for temporal forecasting can be written as follows:
\begin{equation}
\begin{aligned}
    \Bh^{(n)} &= f_h^h(\ba^{(n)}, \Bh^{(n-1)}), \\
    \bc^{(n)} &= f_h^o(\Bh^{(n)}),
\end{aligned} \label{eq:dnn}
\end{equation}
where $\ba^{(n)} :=\ba(t_n) \in \mathbb{R}^R$ is the vector of modal coefficients at time $t_n$ and $\bc^{(n)} \in \mathbb{R}^R$ is the corresponding closure term, defining the input and output of the DNN, respectively. In Eq.~\cref{eq:dnn}, $\Bh \in \mathbb{R}^H$ represents the hidden-state of the neural network, %with 
$f_h^h$ and $f_h^o$ %being 
the hidden-to-hidden and hidden-to-output mappings, respectively, and $H$ %is 
the dimension of the hidden state. The Mori-Zwanzig formulation \cite{stinis2015renormalized,zwanzig2001nonequilibrium,gouasmi2017priori,pan2018data,wang2020recurrent} shows that non-Markovian terms are required to account for the effects of the unresolved scales onto the resolved scales.
%\AV{Curiosity: why non-Markovian? Because it is involving more than just the previous step?}
%\ti{Yes.}
Thus, the closure operators are modeled as functions of the time history of the resolved scales. %Indeed, 
We emphasize that employing a non-Markovian closure model is a key %charachter 
feature of the proposed PGML-VMS-ROM %, as opposed to 
that is in stark contrast with the DD-VMS in \cite{mou2021data,koc2021verifiability}, which considers only the Markovian effects. %only. 
%\ti{I think we should emphasize the non-Markovian character of the PGML-VMS-ROM.  That's one major difference from the D2-VMS-ROM.  I think these differences should be clearly empahsized throughout the paper.}

For memory embedding, we let $\bc$ be a function of the short time history of the resolved POD coefficients, i.e., $\bc^{(n)}(\cdot) \approx \pi_{\theta}(\ba^{(n)},\ba^{(n-1)},\dots, \ba^{(n-\tau)}) = \pi_{\theta}(\ba^{(n):(n-\tau)})$, where $\tau$ defines the length of the time history of $\ba$ that is required for estimating the closure term. The LSTM allows modeling non-Markovian processes while mitigating the issue with vanishing (or exploding) gradient by employing gating mechanisms. In particular, the hidden-to-hidden mapping $f_h^h$ is defined using the following equations:
\begin{equation}
    \begin{aligned}
        \Bg_f^{(n)} &= \sigma_{f}(\BW_f[\Bh^{(n-1)},\ba^{(n)}] + \Bb_f), \\
        \Bg_i^{(n)} &= \sigma_{i}(\BW_i[\Bh^{(n-1)},\ba^{(n)}] + \Bb_i), \\
        \tilde{\Bs}^{(n)} &= \tanh{(\BW_s[\Bh^{(n-1)},\ba^{(n)}] + \Bb_s)}, \\
        \Bs^{(n)} &= \Bg_f^{(n)} \odot \Bs^{(n-1)} + \Bg_i^{(n)} \odot \tilde{\Bs}^{(n)}, \\
        \Bg_o^{(n)} &= \sigma_{o}(\BW_o[\Bh^{(n-1)},\ba^{(n)}] + \Bb_o), \\
        \Bh^{(n)} &= \Bg_o^{(n)} \odot \tanh{(\Bs^{(n)})},
    \end{aligned}
    \label{eqn:ml-vms-2-2}
\end{equation}
where $\Bg_f, \Bg_i, \Bg_o \in \mathbb{R}^H$ are the forget gate, input gate, and output gate, respectively, with the corresponding $\BW_f, \BW_i, \BW_o \in \mathbb{R}^{H \times (H+R)}$ weight matrices, and $\Bb_f, \Bb_i, \Bb_o \in \mathbb{R}^{H}$ bias vectors. $\Bs \in \mathbb{R}^H$ is the cell state with a weight matrix $\BW_s \in \mathbb{R}^{H \times (H+R)}$ and bias vector $\Bb_s \in \mathbb{R}^H$. Finally,  $\sigma$ is the sigmoid activation function, and $\odot$ denotes the element-wise multiplication.

We stack $l$ LSTM layers to define the hidden states, followed by a fully connected layer with a linear activation function to represent the hidden-to-output mapping. Thus, the ML-VMS-2 closure model can be written as
\begin{equation}
\begin{aligned}
        \bc^{(n)} \approx \mathcal{L} (\cdot) \circ \Bh_l^{(n)}(\cdot) \circ \Bh_{l-1}^{(n):(n-\tau)}(\cdot) \circ \dots \circ \Bh_1^{(n):(n-\tau)}(\cdot)  \circ \mathcal{I} (\ba^{(n):(n-\tau)}),\\
\end{aligned}
    \label{eqn:ml-vms-2-3}
\end{equation}
where $\mathcal{L} (\cdot)$ represents the output layer with linear activation, and $\mathcal{I} (\cdot)$ denotes the input layer. Note that each of the internal LSTM layers ($i=1,2,\dots, l-1$) produces a sequence of hidden states $\Bh_i^{(n):(n-\tau)}$, while the the $l^{th}$ layer passes only the hidden state at the final time $\Bh_l^{(n)}$ to the output layer. 

To summarize, Eqs.~\eqref{eqn:ml-vms-2-1}, \eqref{eq:dnn}, \eqref{eqn:ml-vms-2-2}, and \eqref{eqn:ml-vms-2-3} yield the ML-VMS-2 ROM.
%The LSTM training amounts to optimizing the weights and biases using the backpropagation through time (BPTT) algorithm, with the following cost function: %In the present study, we utilize the LSTM neural network to represent the closure model, i.e.,  $\bc_k(\cdot) \approx \pi_{\theta}(\ba_k,\ba_{k-1},\dots, \ba_{k-\tau})$, where $\theta$ denotes the parameterization of the LSTM.
In order to make use of the locality of %energy transfer, 
modal interactions, the VMS-3 ROM is written as%:
\begin{equation}
    \begin{bmatrix} \dot{\ba_L} \\ \dot{\ba_S}  \end{bmatrix} = 
    \bff(\ba) +
    \begin{bmatrix} \bc_L(\ba) \\ \bc_S(\ba)  \end{bmatrix},
\end{equation}
where two separate %models 
terms are dedicated to model the closure for the resolved large scales and resolved small scales. For the ML-VMS-3, the closure terms are defined as follows:
\begin{equation}
\begin{aligned}
        \bc^{(n)}_{L} &\approx \pi_{L,\theta}(\ba^{(n):(n-\tau)}) \\
         & \approx \mathcal{L}_{_L} (\cdot) \circ \Bh_{l_L}^{(n)}(\cdot) \circ \Bh_{l-1_L}^{(n):(n-\tau)}(\cdot) \circ \dots \circ \Bh_{1_L}^{(n):(n-\tau)}(\cdot)  \circ \mathcal{I} (\ba^{(n):(n-\tau)}),\\
%%%%%%%%%%%%%%%%%%%%%%%
        \bc^{(n)}_{S} &\approx \pi_{S,\theta}(\ba^{(n):(n-\tau)}) \\
         & \approx \mathcal{L}_{_S} (\cdot) \circ \Bh_{l_S}^{(n)}(\cdot) \circ \Bh_{l-1_S}^{(n):(n-\tau)}(\cdot) \circ \dots \circ \Bh_{1_S}^{(n):(n-\tau)}(\cdot)  \circ \mathcal{I} (\ba^{(n):(n-\tau)}).
\end{aligned}
\end{equation}
We note that we have more flexibility in ML-VMS-3 than in ML-VMS-2. Hence, it is possible to make richer descriptions %on 
of the interactions between large resolved, small resolved, and unresolved scales.

 %\AV{The previous description does not clarify the substantial difference 
 %between the two terms. Where can we infer that ones describes the interaction large resolved/small resolved and the other the interaction small resolved/unresolved? I think we should say something more here.}
 %\ti{I agree.  Maybe just say that we have more flexibility in ML-VMS-3?}

\subsection{PGML-VMS ROM}
Critical aspects that should be considered during the adoption of %machine learning (ML) 
ML based approach include their reliability, robustness, and trustworthiness. Previous studies~\cite{pawar2021physics,pawar2021model,pawar2021multi} have reported high levels of uncertainty in the predictions of vanilla-type ML methods, especially for sparse data and incomplete governing equations regimes. In order to mitigate this issue, we utilize the physics-guided machine learning (PGML) paradigm to incorporate %the 
known physical arguments and constraints into the learning process. In particular, we exploit a modular approach to modify the neural network architectures through layer concatenation to inject physical information at different points in the latent space of the given DNN. This adaptation augments the performance during both the training and the deployment phases, and results in significant reduction in the uncertainty levels of the model prediction, as we demonstrate in~\cref{sec:res}. 

In the PGML framework, the features extracted from the physics-based model are embedded into the generic $i^{th}$ intermediate hidden layer along with the latent variables. In order to build the PGML-VMS framework, we consider the Galerkin projection of the governing equations onto different POD modes to define the physics-based features (since they are derived from 
physical principles).
Thus, the PGML-VMS-2 closure model can be written as
\begin{equation}
\begin{aligned}
        \bc^{(n)} \approx \mathcal{L} (\cdot) \circ \Bh_l^{(n)}(\cdot) &\circ \dots \circ \mathcal{C}\bigg(\Bh_i^{(n):(n-\tau)}(\cdot), \bff^{(n):(n-\tau)} \bigg) \circ \Bh_{i-1}^{(n):(n-\tau)}(\cdot) \circ  \dots \circ \Bh_1^{(n):(n-\tau)}(\cdot)  \circ \mathcal{I} (\ba^{(n):(n-\tau)}),
\end{aligned} \label{eq:pgml2}
\end{equation}
where $\mathcal{C}(\cdot,\cdot)$ represents the concatenation operation, and $\bff^{(n):(n-\tau)}$ is the time history of projecting the FOM operators onto the truncated POD subspace. We highlight that there is no significant computational load for the calculation of  $\bff := A\ba + \ba^\top B \ba$, since $A$ and $B$ are already precomputed.
%\ti{Is this projection expensive?} \SA{here $\bff := A\ba + \ba^\top B \ba$. Part of it ($A$ and $B$) is precomputed, and the remaining part (e.g., matrix-vector multiplications) is inevitable anyway for GROM - so no extra computations here}

\begin{figure}[h]
\begin{center}
	\includegraphics[width=0.9\textwidth]{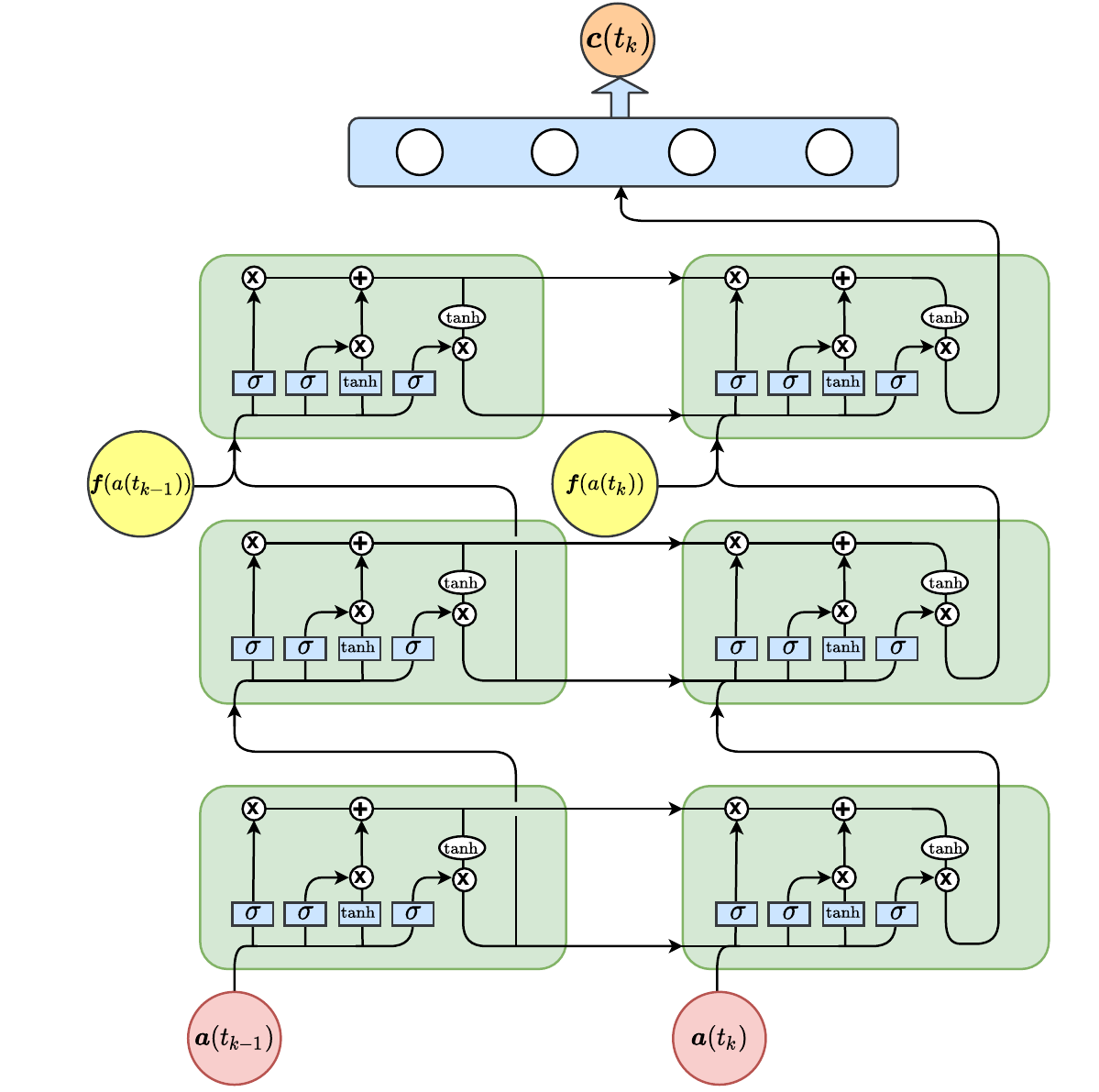}
    \caption{Illustration of the PGML methodology with concatenated LSTM layers. In this figure, a time history of $2$ time steps is used while physics-based features (yellow circles in the figure) are injected into the LSTM latent space after the second hidden layer ($i=2$).}
	\label{fig:pgml}
\end{center}
\end{figure}

A schematic illustration of the PGML adaptation of the standard LSTM architecture is depicted in~\cref{fig:pgml}. In this figure, 3 LSTM layers are used (i.e., $l=3$), followed by a dense layer to provide the mapping from hidden state to the closure terms. The physics-based features are injected into the LSTM latent space after two hidden layers.  One of the main advantages of the novel PGML framework in~\cref{fig:pgml} is its modularity and simplicity. For example, based on the level of fidelity and our confidence in the injected features, we can promptly change the layer at which we embed them.

Finally, the PGML-VMS-3 closure models can be written as
\begin{equation}
\begin{aligned}
        \bc^{(n)}_L \approx \mathcal{L}_{_L} (\cdot) \circ \Bh_{l_L}^{(n)}(\cdot) &\circ \dots \circ \mathcal{C}\bigg(\Bh_{i_L}^{(n):(n-\tau)}(\cdot), \bff_{_L}^{(n):(n-\tau)} \bigg) \circ \Bh_{i-1_L}^{(n):(n-\tau)}(\cdot) \circ  \dots \circ \Bh_{1_L}^{(n):(n-\tau)}(\cdot)  \circ \mathcal{I} (\ba^{(n):(n-\tau)}),\\
        %%%%%%%%%%%%%%%%%%%%%%%%%%%%%%%%%%%
        \bc^{(n)}_S \approx \mathcal{L}_{_S} (\cdot) \circ \Bh_{l_S}^{(n)}(\cdot) &\circ \dots \circ \mathcal{C}\bigg(\Bh_{i_S}^{(n):(n-\tau)}(\cdot), \bff_{_S}^{(n):(n-\tau)} \bigg) \circ \Bh_{i-1_S}^{(n):(n-\tau)}(\cdot) \circ  \dots \circ \Bh_{1_S}^{(n):(n-\tau)}(\cdot)  \circ \mathcal{I} (\ba^{(n):(n-\tau)}).
\end{aligned} \label{eq:pgml3}
\end{equation}
Note that in Eq.~\cref{eq:pgml3}, we enjoy higher flexibility in choosing the physics-based features injected for each of the large and small scale closure models. For instance, in the present study, we benefit from the locality of %energy transfer 
modal interactions by embedding the Galerkin propagator of only a few relevant neighboring modes (i.e., $\bff_{_L}$ and $\bff_{_S}$ in \cref{eq:pgml3}), rather than including all of them in the LSTM learning (i.e., $\bff$ in \cref{eq:pgml2}).
% \AV{Is the previous statement somehow reflected in the notation of \ref{eq:pgml3}? If yes, we should stress it.}
% \ti{I agree.}

%%========================================%%
\section{Nonlinear POD} \label{sec:nlpod}
\begin{figure}[h]
\begin{center}
	\includegraphics[width=1.0\textwidth]{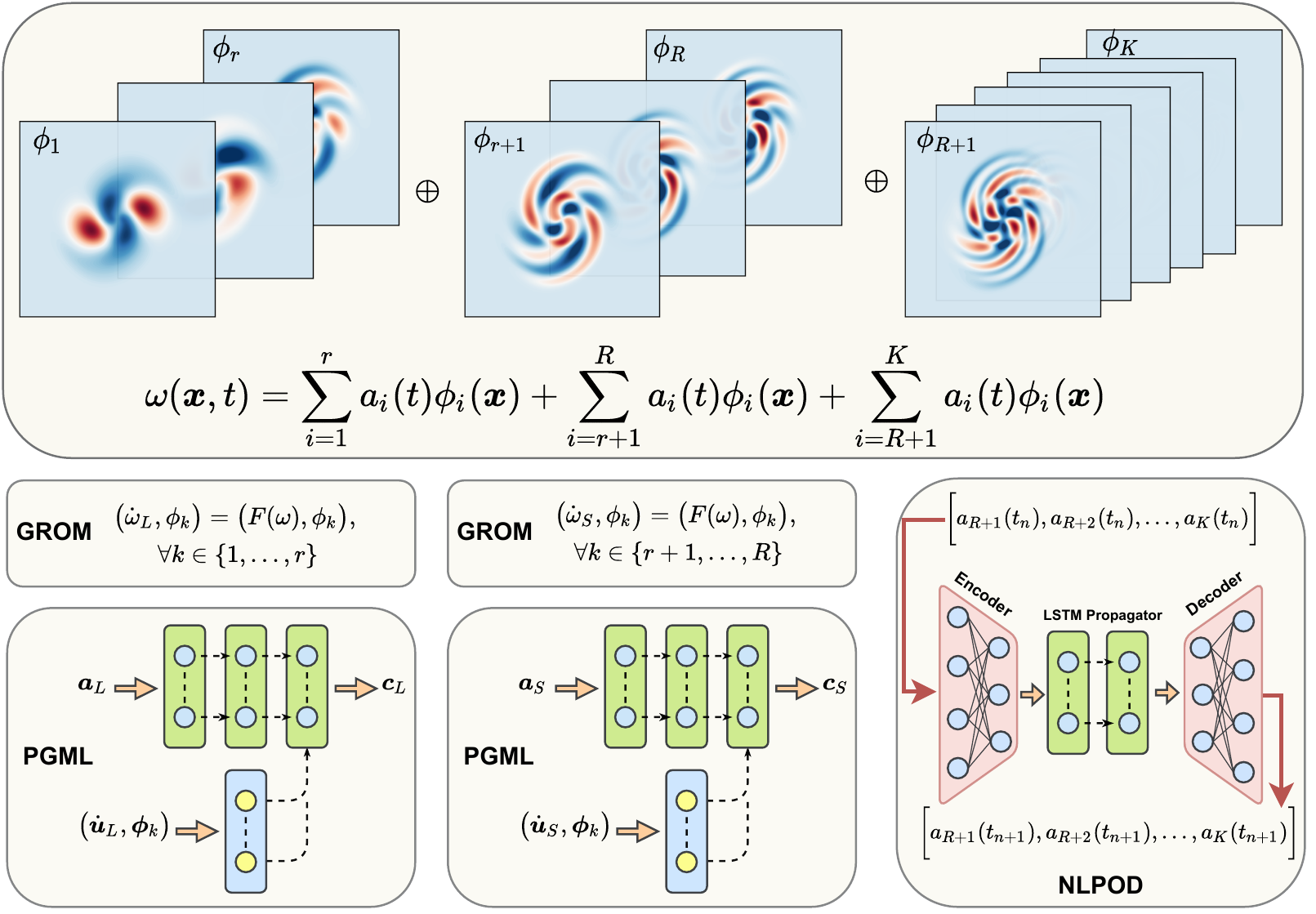}
    \caption{Schematic representation of the %combined 
    PGML-VMS-3 model for the large and small resolved scales, %along 
    combined with NLPOD for enhanced field reconstruction. 
    %\ti{accuracy}.
    We note that PGML-VMS-3 is built upon a GROM for the first $R$ modes and %addresses 
    mitigates the closure error (i.e., the effect of the truncated scales onto the resolved scales). In a complementary %yet separate 
    fashion, NLPOD implements an equation-free model for the truncated scales to %eventually 
    reduce the projection error (i.e., the effect of the truncated scales onto the flow field reconstruction).}
	\label{fig:framework}
\end{center}
\end{figure}

%%========================================%%
In~\cref{sec:vms} and~\cref{sec:pgml}, we addressed the closure problem. That is, we %are interested in 
aimed at correcting the ROM equations for the dynamics of the resolved scales including %We model 
%Specifically, we modeled 
the effects of the unresolved scales onto the dynamics of the resolved scales. %However, eventually 
However, % that 
the reconstructed flow fields %are 
were approximated within the span of the retained modes, %see 
as shown in Eq.~\cref{eq:upod}. Nonetheless, for turbulent flows %and convection-dominated flow systems, 
the important flow structures generally span a large number of modes. %Truncating 
Thus, truncating the solution beyond a small number of modes results in a large projection error. In other words, the component $\omega' = \sum_{k=R+1}^{N} a_k \phi_k$ that %lives in the null space of 
cannot be approximated by 
the resolved POD basis becomes significant. In this section, we adapt the nonlinear POD (NLPOD) framework, introduced in~\cite{ahmed2021nonlinear}, to model the unresolved part of the field.
\cref{fig:framework} presents a schematic representation of the %combined 
PGML-VMS-3 model for the large and small resolved scales %along 
combined with NLPOD for enhanced field reconstruction. %in order to illustrate how the NLPOD component is different from the PGML component.
Note that, although both the PGML-VMS-3 and the NLPOD aim at increasing the ROM accuracy, they target different error sources:
the PGML-VMS-3 aims at mitigating the closure error, whereas the NLPOD aims at alleviating the projection error.

%\ti{I agree with Ale's email.  I think we need to better explain how the NLPOD component is different from the PGML component.  For example, the effect of $\bu'$ is modeled by closure.  How is NLPOD different?  Maybe adding a schematic/illustration with the differences between the two components?}

The NLPOD methodology is based on combining POD with autoencoder (AE) techniques from ML to learn a latent representation of the POD expansion. It leverages the predefined hierarchy of POD basis functions, which satisfy the conservation laws and physical constraints, together with the capabilities of DNN %in revealing 
to reveal the nonlinear correlations between the modes. Rather than using the NLPOD for the compression of the total set of POD coefficients, we constrain it to learn a few latent variables, %that only 
which represent only the unresolved scales. %In particular, we 
To construct the NLPOD, we first    
define $\bb = \{a_k\}_{k=R+1}^{K}$ corresponding to an almost full-rank POD expansion, where $K\le N$ can be defined using the RIC spectrum (e.g., $\text{RIC}(K) \ge 99.99\%$). %We 
The goal is to learn $\bz = \{z_k\}_{k=1}^{q}$, where $q \ll K$ denotes the dimension of the AE latent space. 

The AE starts with an encoding process that involves applying a series of nonlinear mappings onto the input data to shrink the dimensionality down to a bottleneck layer representing the low rank or latent space embedding. An inverse mapping from the latent space variables to the same input is performed by another set of nonlinear mappings, defining the decoding part. For the NLPOD, the encoder and decoder can be represented as follows:
\begin{equation}
    \begin{aligned}
            \text{Encoder} \ \eta&: \bb \in \mathbb{R}^{K-R} \mapsto \bz\in \mathbb{R}^{q}, \qquad
            \text{Decoder} \ \zeta&: \bz\in \mathbb{R}^{q}  \mapsto 
            \bb\in \mathbb{R}^{K-R} ,
    \end{aligned}
\end{equation}
and they are trained jointly to minimize the following objective function:
\begin{equation}
    \mathcal{J} = \sum_{n=1}^{N_{train}}\| \bb^{(n)} - (\eta \circ \zeta)(\bb^{(n)}) \|,
\end{equation}
where $N_{train}$ is the number of training samples. 

In order to temporally propagate %the 
$\bz$, we can use any of the regression tools, including sparse regression, Gaussian process regression, Seq2seq algorithms, temporal fusion transformers, and auto-regression methods. In the present study, we use %similar 
LSTM architectures that are similar to the ones used in~\cref{sec:pgml} to learn the one time-step mapping from $\bz^{(n)}$ to $\bz^{(n+1)}$, as follows:
\begin{equation}
\begin{aligned}
        \bz^{(n+1)} \approx \mathcal{L} (\cdot) \circ \Bh_l^{(n)}(\cdot) \circ \Bh_{l-1}^{(n):(n-\tau)}(\cdot) \circ \dots \circ \Bh_1^{(n):(n-\tau)}(\cdot)  \circ \mathcal{I} (\bz^{(n):(n-\tau)}).
\end{aligned}
\end{equation}
Note that the number of layers, $l$, and the length of time history, $\tau$, are not necessarily equal to those in~\cref{sec:pgml}. Moreover, the LSTM and AE can be trained either jointly or separately. In the present study, we train them separately for the sake of simplicity and to facilitate the NLPOD combination with other time series prediction tools.

%[trim={25cm 0 25cm 0},clip, width=0.9\textwidth]
\clearpage
%%================================================%%
\section{Results and Discussion} \label{sec:res}
%%================================================%%
%We demonstrate the developed 
In this section, we perform a numerical investigation of the proposed PGML-VMS-ROM
%\ti{(Omer, I think we should be consistent when using these acronyms.  Maybe we can also add a little table where we list the acronyms and the corresponding equations?)} \av{GOOD IDEA, AS FOR NOW I PLACED HERE THE ACRONYMS USED IN THE INTRO, BUT I NOTICED THAT IN THE FOLLOWUP YOU DON'T USE THEM} 
methodologies (with and without the NLPOD extension) using the two dimensional (2D) vortex merger problem \cite{san2013coarse}, governed by the following vorticity transport equation:
\begin{equation}
\partial_t \omega + J(\omega,\psi) = \dfrac{1}{\text{Re}} \Delta \omega, \qquad \text{in} \ \Omega \times [0,T]. 
\end{equation}
% where $\omega$ and $\psi$ denote the vorticity and streamfunction fields, and ($J(\cdot,\cdot)$) is the Jacobian operator defined as:
% \begin{align}
%     J(\omega,\psi) &= \dfrac{\partial \omega}{\partial x} \dfrac{\partial \psi}{\partial y} -  \dfrac{\partial \omega}{\partial y} \dfrac{\partial \psi}{\partial x}.
% \end{align}
% The vorticity and stramfunction are linked by the kinematic relationship:
% \begin{equation}
% \Delta  \psi = -\omega. \label{eq:Poisson}
% \end{equation}

\begin{figure}[h]
\begin{center}
	\includegraphics[width=0.8\textwidth]{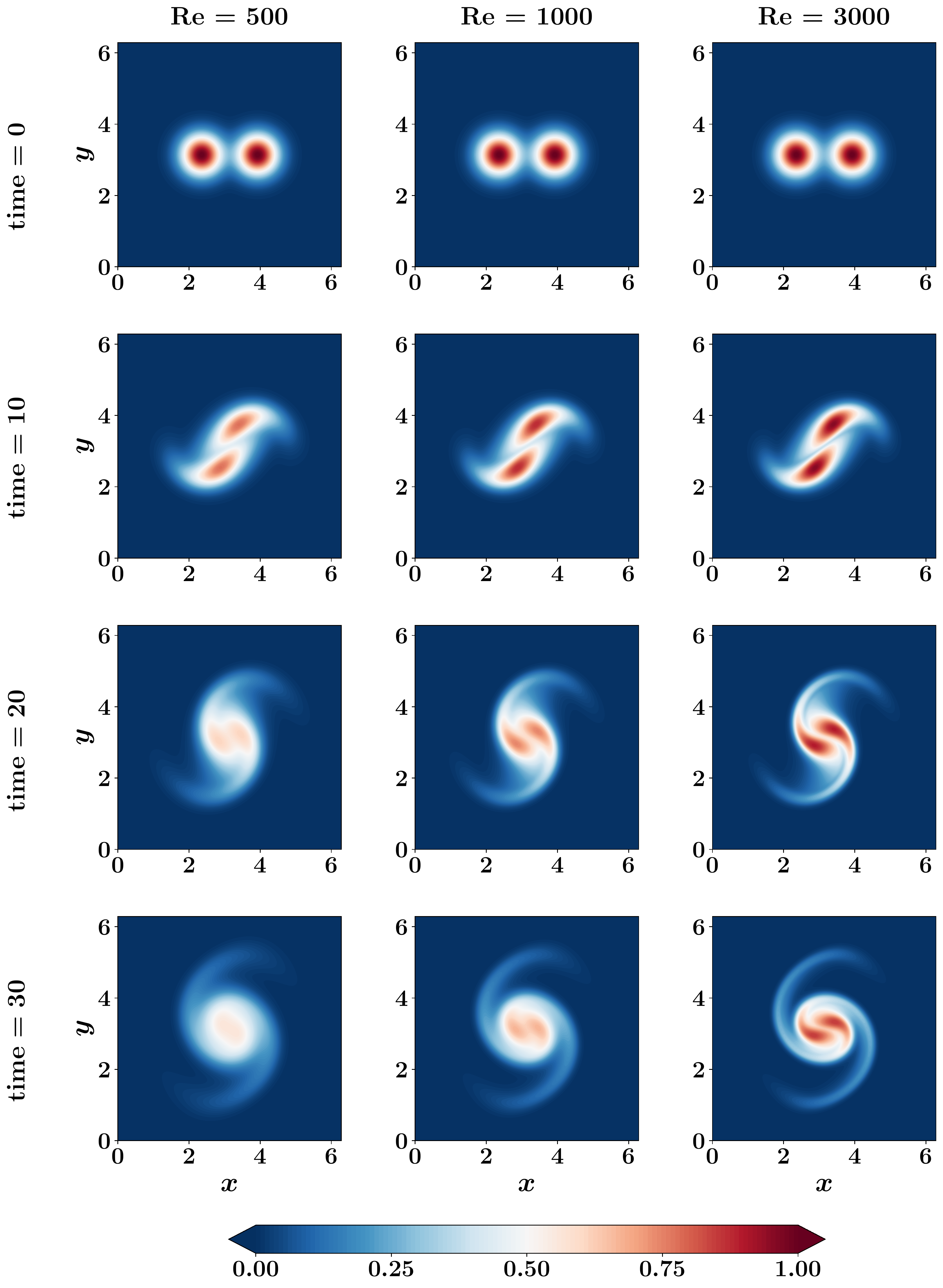}
    \caption{Samples of temporal snapshots of the vorticity field for the vortex merger problem at different values of Reynolds number.}
	\label{fig:fom}
\end{center}
\end{figure}

We consider a spatial domain of dimensions $(2\pi \times 2\pi)$ with periodic boundary conditions. The flow is initialized with a pair of co-rotating Gaussian vortices with equal strengths centered at $(x_1,y_1) = ( 5\pi/4,\pi)$ and $(x_2,y_2) = ( 3\pi/4,\pi)$ as follows:
\begin{equation}
    \omega(x,y,0) =  \exp\left( -\rho \left[ (x-x_1)^2  + (y-y_1)^2 \right] \right) + \exp{\left( -\rho \left[ (x-x_2)^2 + (y-y_2)^2 \right] \right)}, \label{eq:init}
\end{equation}
where $\rho$ is %an interacting 
a parameter that controls the mutual interactions between the two vortical motions, set at $\rho = \pi$ in the present study. For the FOM simulations, we consider a regular Cartesian grid resolution of $256\times256$ (i.e., $\Delta x = \Delta y = 2\pi/256$), with a time-step of $0.001$. %Snapshots of vorticity fields 
Vorticity snapshots 
are collected every 100 time-steps for $t\in [0,30]$, totalling 300 snapshots. The evolution of the vortex merger problem at selected values of the Reynolds number is depicted in~\cref{fig:fom}, %demonstrating 
which illustrates 
the convective and interactive mechanisms affecting the transport and development of the two vortices.

\begin{figure}[h]
\begin{center}
	\includegraphics[width=0.61\textwidth]{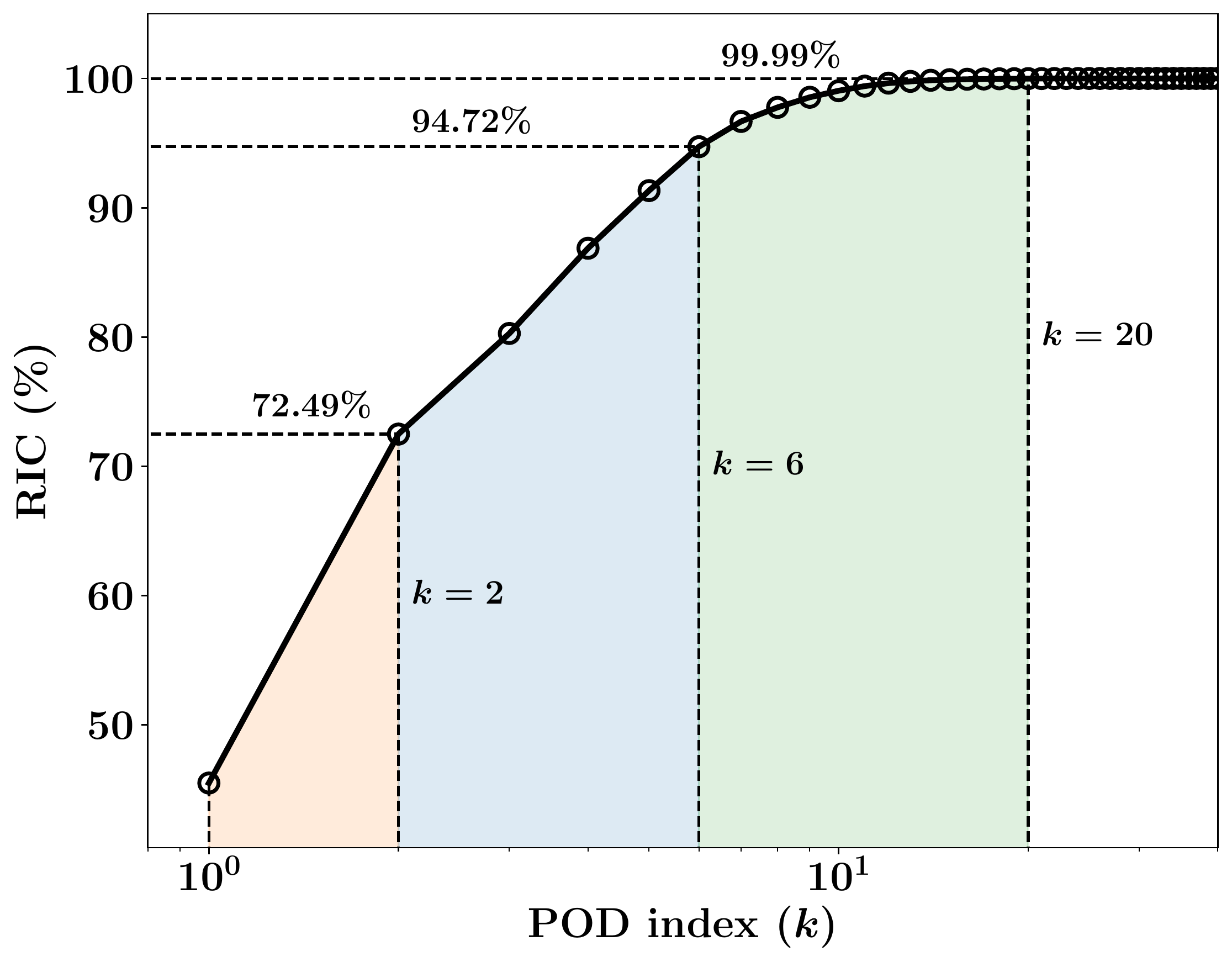}
    \caption{RIC values %at different 
    as a function of the modal truncation for the vortex merger problem at $\text{Re} = 3000$.} %\textcolor{blue}{Shady: $k=2$ corresponds to the first level in VMS (i.e., the large resolved scales), $k=6$ corresponds to the second level (the small resolved scales), and $k=20$ denotes the cut-off at $\text{RIC}=99.99\%$ for NLPOD -- sort of small unresolved scales.}}
	\label{fig:ric}
\end{center}
\end{figure}

In terms of POD analysis, we use $R=6$ to define the total number of resolved scales. For the three-scale VMS %demonstration, 
investigation, 
we split the resolved modes into 2 resolved large scales (i.e., $r=2$) and 4 resolved small scales. For the NLPOD study, we find that $K=20$ corresponds to near full-rank approximation of the flow field at all values of the Reynolds number. 
This is illustrated by the plot of the RIC values as a function of the number of POD modes at $\text{Re}=3000$ %are plotted 
in~\cref{fig:ric}.

Following a systematic approach, in~\cref{sec:6.1}, we first present our computational results for ML-VMS-2 and PGML-VMS-2 to quantitatively demonstrate the benefit of incorporating the physics guided machine learning approach. We then present
the  results for PGML-VMS-3 to highlight the flexibility and accuracy gain of the three-scale approach. Finally, in~\cref{sec:6.2}, we reveal the additional role of the NLPOD approach by illustrating the performance of the PGML-VMS-3+NLPOD approach.  

\subsection{Multi-level VMS closure for resolved scales} \label{sec:6.1}
\begin{figure}[h]
\begin{center}
	\includegraphics[width=0.83\textwidth]{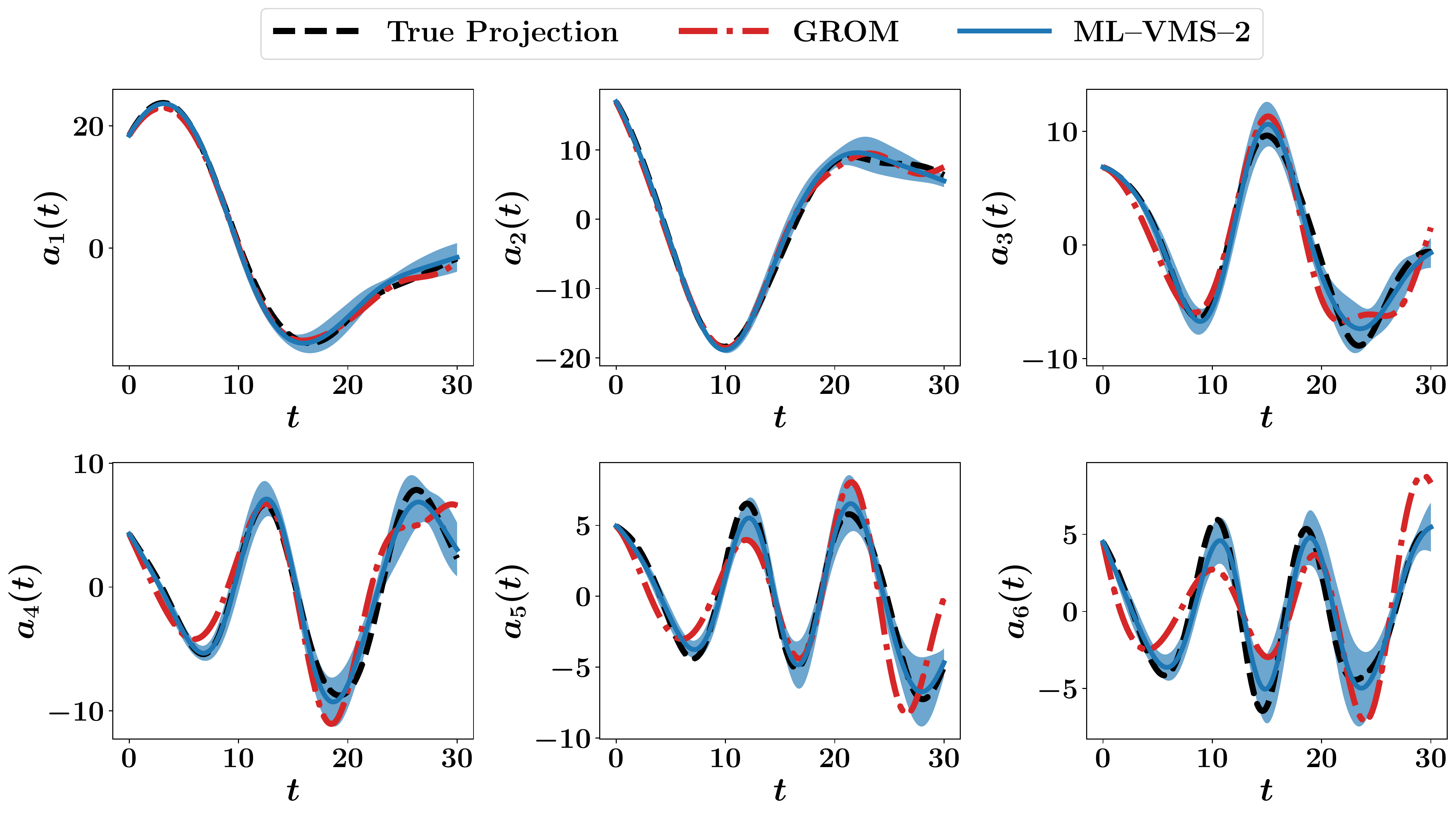}
    \caption{The time evolution of the first 6 modes of the vortex merger problem with the two-level VMS using ML closure, compared to the true projection and GROM (without closure) predictions. The solid line represents the mean values ($\mu$) from an ensemble of 10 different LSTM neural networks trained with different weight initalizations, while the shaded area defines the uncertainty bounds using standard deviation ($\sigma$) values. For better visualization, the shaded band is plotted with $\mu \pm 5\sigma$.}
	\label{fig:ml2}
\end{center}
\end{figure}

We store data corresponding to $\text{Re} \in \{500,750,1000,\dots,3000\}$ (in increments of 250), but we use only the data collected at $\text{Re} \in \{500,750,1000\}$ for neural network training, while the %rest 
remaining data is reserved for testing purposes. First, we explore the combination of multi-level variational multiscale methods with machine learning. \cref{fig:ml2} displays the results of applying the ML-VMS-2 framework to model the closure term at $\text{Re}=3000$. In particular, we run a group of 10 LSTMs with different initializations of the neural network weights and utilize the deep ensemble method to quantify the uncertainty in the predictions. %We can see that the 
On the average, the ML-VMS-2 method provides accurate %ROMs 
results compared to the GROM results. However, the uncertainty levels, described by the standard deviation in the ensemble predictions, are high. This is especially evident at the late time instants as the uncertainty propagates and grows with time.

\begin{figure}[h]
\begin{center}
	\includegraphics[width=0.83\textwidth]{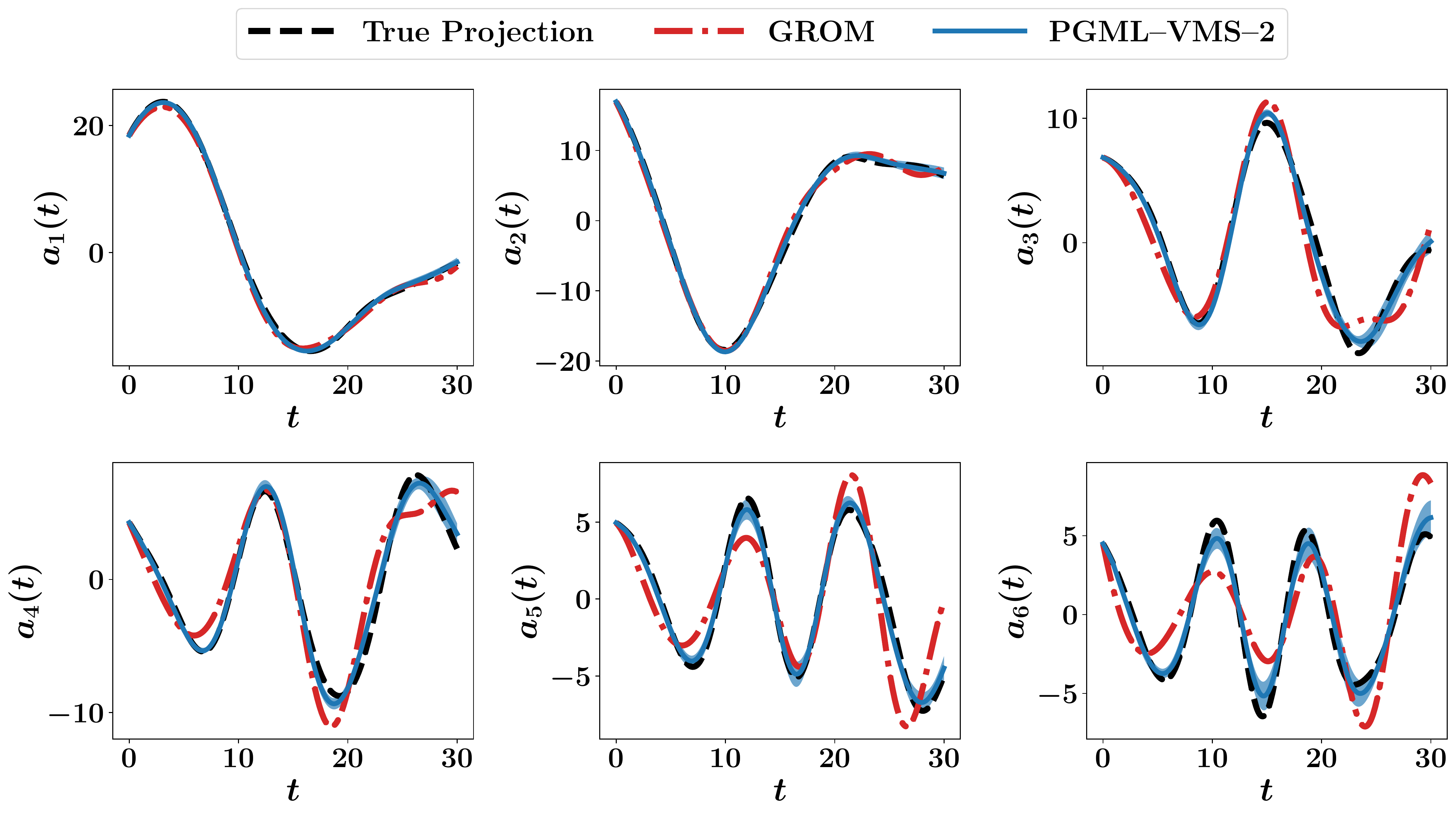}
    \caption{The time evolution of the first 6 modes of the vortex merger problem with the two-level VMS using PGML closure, compared to the true projection and GROM (without closure) predictions. The solid line represents the mean values ($\mu$) from an ensemble of 10 different LSTM neural networks trained with different weight initalizations, while the shaded area defines the uncertainty bounds using standard deviation ($\sigma$) values. For better visualization, the shaded band is plotted with $\mu \pm 5\sigma$.}
	\label{fig:pgml2}
\end{center}
\end{figure}

In order to increase the closure model robustness and reduce the uncertainty levels, we apply the PGML to inject physics-based features,  as detailed in~\cref{sec:pgml}. \cref{fig:pgml2} shows the evolution of the first 6 POD modal coefficients using the PGML-VMS-2. We can observe a significant reduction in the uncertainty levels as depicted by the shaded area, compared to the ML-VMS-2. %We note that the GROM results correspond to applying the Galerkin method without correction (refer to Eq.~\cref{eq:GNSE} and Eq.~\cref{eq:Gtens}). 
It is also clear that the GROM yields inaccurate predictions. %by time. 
Moreover, we can observe that the deviations of the GROM trajectory from the true projections are larger for the latest resolved modes. In fact, this observation also applies to the ML-VMS-2 and PGML-VMS-2, which provide better results for the first two or three modes than the remaining ones.

\begin{figure}[h]
\begin{center}
	\includegraphics[width=0.83\textwidth]{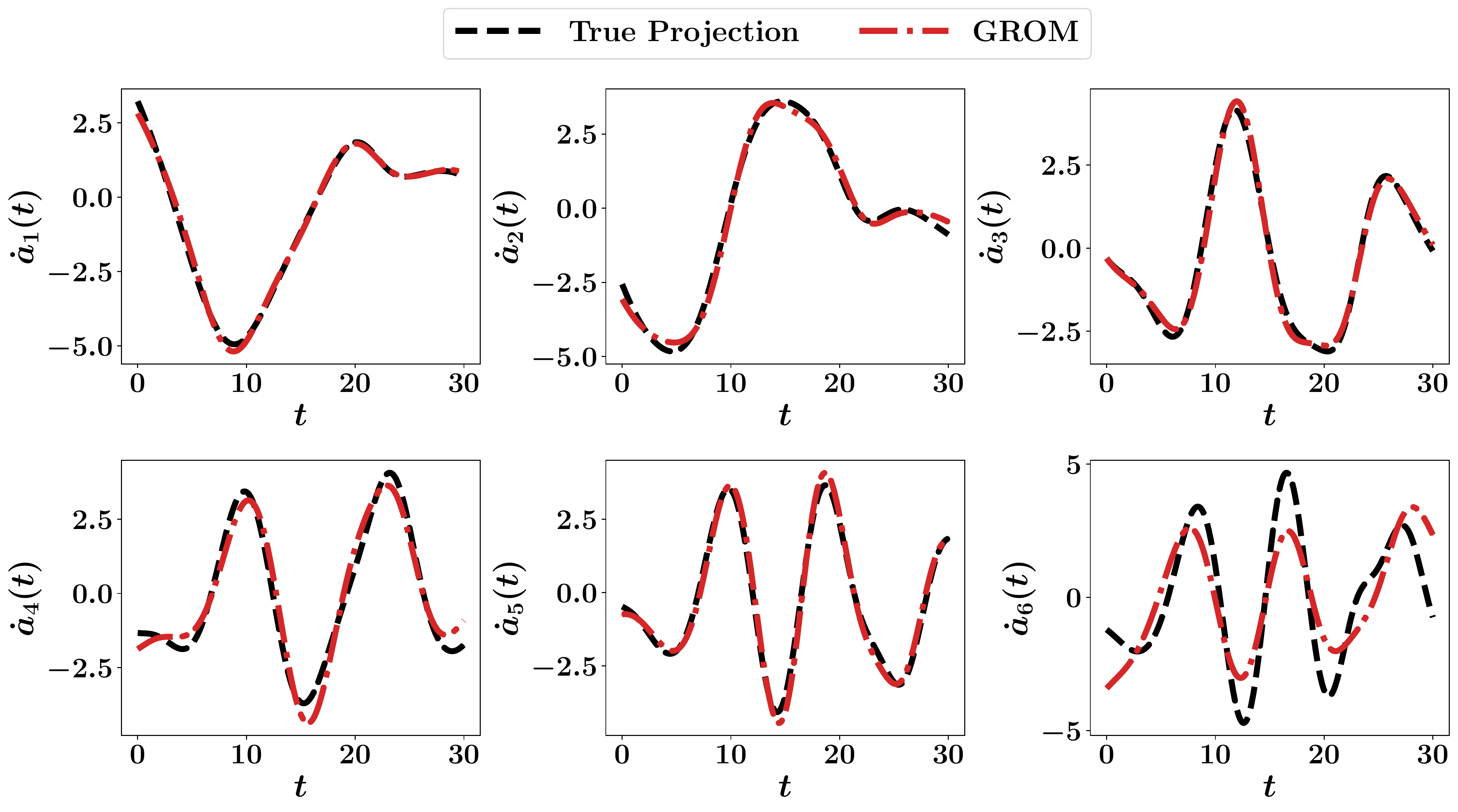}
    \caption{Comparison between the ROM propagator computed by Galerkin projection (with truncation, i.e., $\dot{a_k} = ( -J(\omega_R,\psi_R) + \nabla^2 \omega_R , \phi_k$), against the true (FOM projection) propagator (i.e., $\dot{a_k} = ( -J(\omega,\psi) + \nabla^2 \omega , \phi_k$) at $\text{Re} = 3000$, and $R=6$. We notice that the Galerkin projection accurately captures the dynamics of the first modes, but a discrepancy appears at the latest modes, which motivates the use of multi-level VMS closure.}
	\label{fig:rhs}
	% $\dot{a_k} = \mathcal{C}_k + \sum_{i=1}^{R} \mathcal{L}_{ik} a_i + \sum_{i=1}^{R} \sum_{j=1}^{R} \mathcal{N}_{ijk} a_i a_j$
\end{center}
\end{figure}

In~\cref{fig:rhs}, we plot the ROM propagator $\dot{\ba}$ computed by the Galerkin method (i.e., with truncation, with no access to the unresolved scales, and without correction) against the true propagator (assuming access to all the flow scales). We find that the GROM equations can adequately describe the dynamics of the first modes, but %fails 
fail to do so for the last ones. This can be explained by locality of information %energy 
transfer, which is one of the main %arguments of the VMS approach. 
concepts used in the VMS development. 
Such locality indicates that the neighboring modes exhibit larger mutual interactions than the modes which are far apart. Thus, describing the dynamics of the leading modes requires more information from the first few scales than from the remaining scales. In other words, the resolved scales become almost sufficient to define the propagator of the leading modes. On the other hand, the %later 
last modes are adjacent to the unresolved scales.%, and truncating these scales considerably affects the dynamics of the later modes.
Thus, the mode truncation considerably affects the dynamics of the last modes.

\begin{figure}[h]
\begin{center}
	\includegraphics[width=0.83\textwidth]{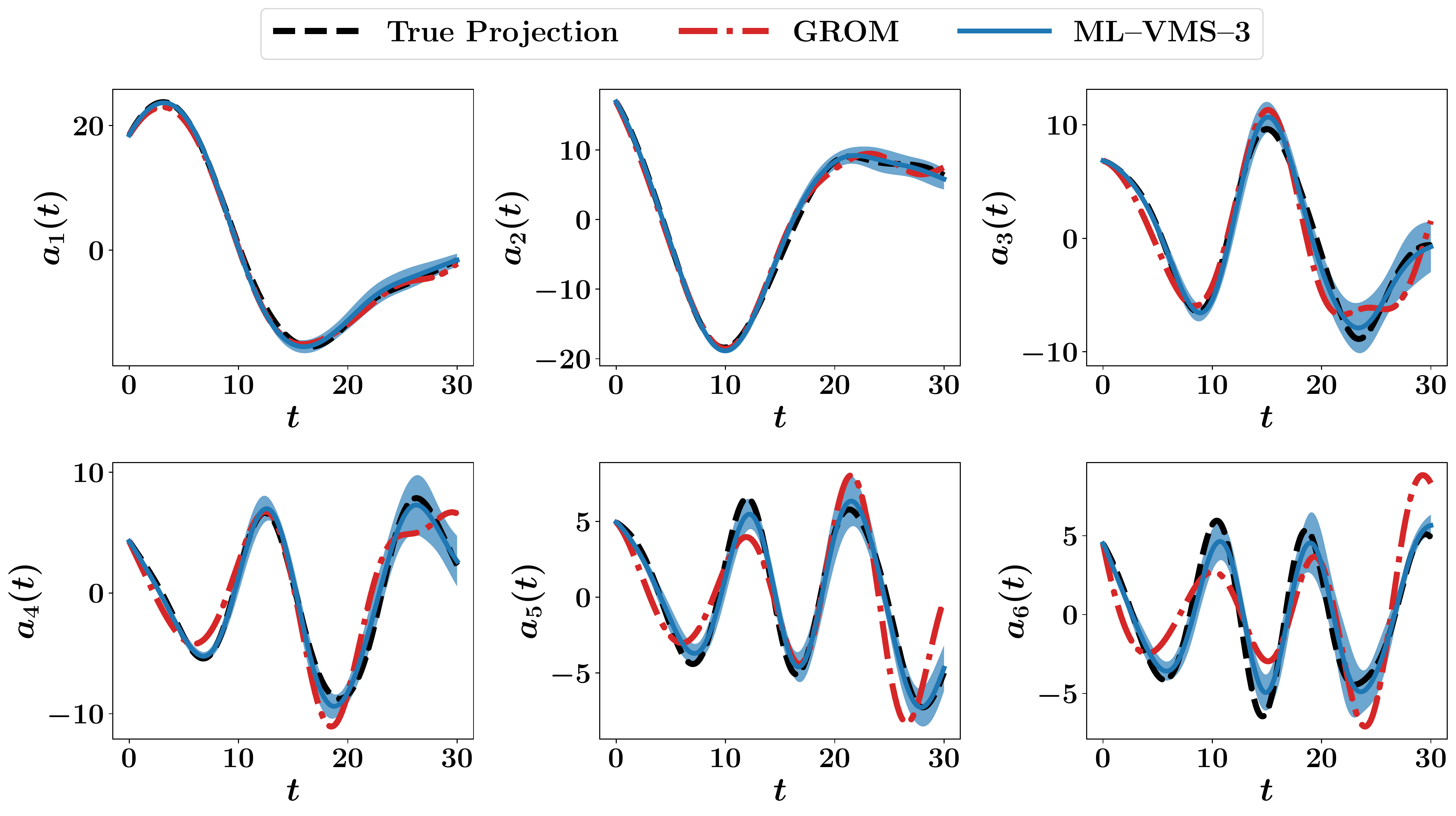}
    \caption{The time evolution of the first 6 modes of the vortex merger problem with the three-level VMS using ML closure, compared to the true projection and GROM (without closure) predictions. The solid line represents the mean values ($\mu$) from an ensemble of 10 different LSTM neural networks trained with different weight initalizations, while the shaded area defines the uncertainty bounds using standard deviation ($\sigma$) values. For better visualization, the shaded band is plotted with $\mu \pm 5\sigma$.}
	\label{fig:ml3}
\end{center}
\end{figure}

In order to improve the quality of the closure model, %and benefit from the locality of energy transfer, 
we leverage the locality of modal interactions and 
%we 
apply the three-level VMS closure to correct the ROM dynamics. In particular, we split the resolved scales into two parts: the first %two 
2 modes represent the largest resolved scales, while the remaining 4 modes represent the small resolved scales. The ML-VMS-3 predictions of the temporal dynamics for the first 6 modes are shown in~\cref{fig:ml3}. Compared to~\cref{fig:ml2}, the ML-VMS-3 provides more accurate results than the ML-VMS-2, even in terms of uncertainty levels.

\begin{figure}[]
\begin{center}
	\includegraphics[width=0.83\textwidth]{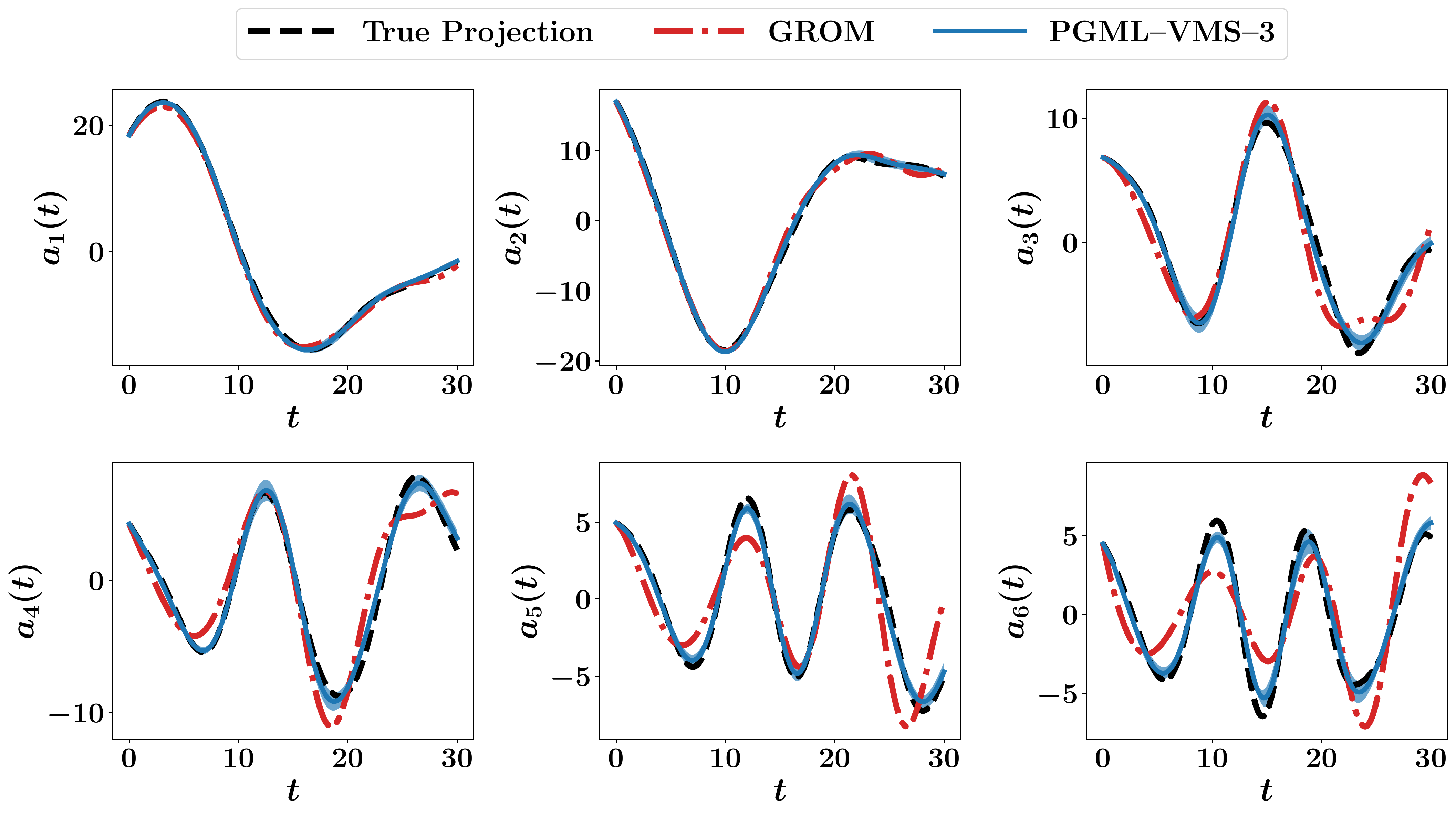}
    \caption{The time evolution of the first 6 modes of the vortex merger problem with the three-level VMS using PGML closure, compared to the true projection and GROM (without closure) predictions. The solid line represents the mean values ($\mu$) from an ensemble of 10 different LSTM neural networks trained with different weight initalizations, while the shaded area defines the uncertainty bounds using standard deviation ($\sigma$) values. For better visualization, the shaded band is plotted with $\mu \pm 5\sigma$.}
	\label{fig:pgml3}
\end{center}
\end{figure}

\begin{figure}[h]
\begin{center}
	\includegraphics[width=0.75\textwidth]{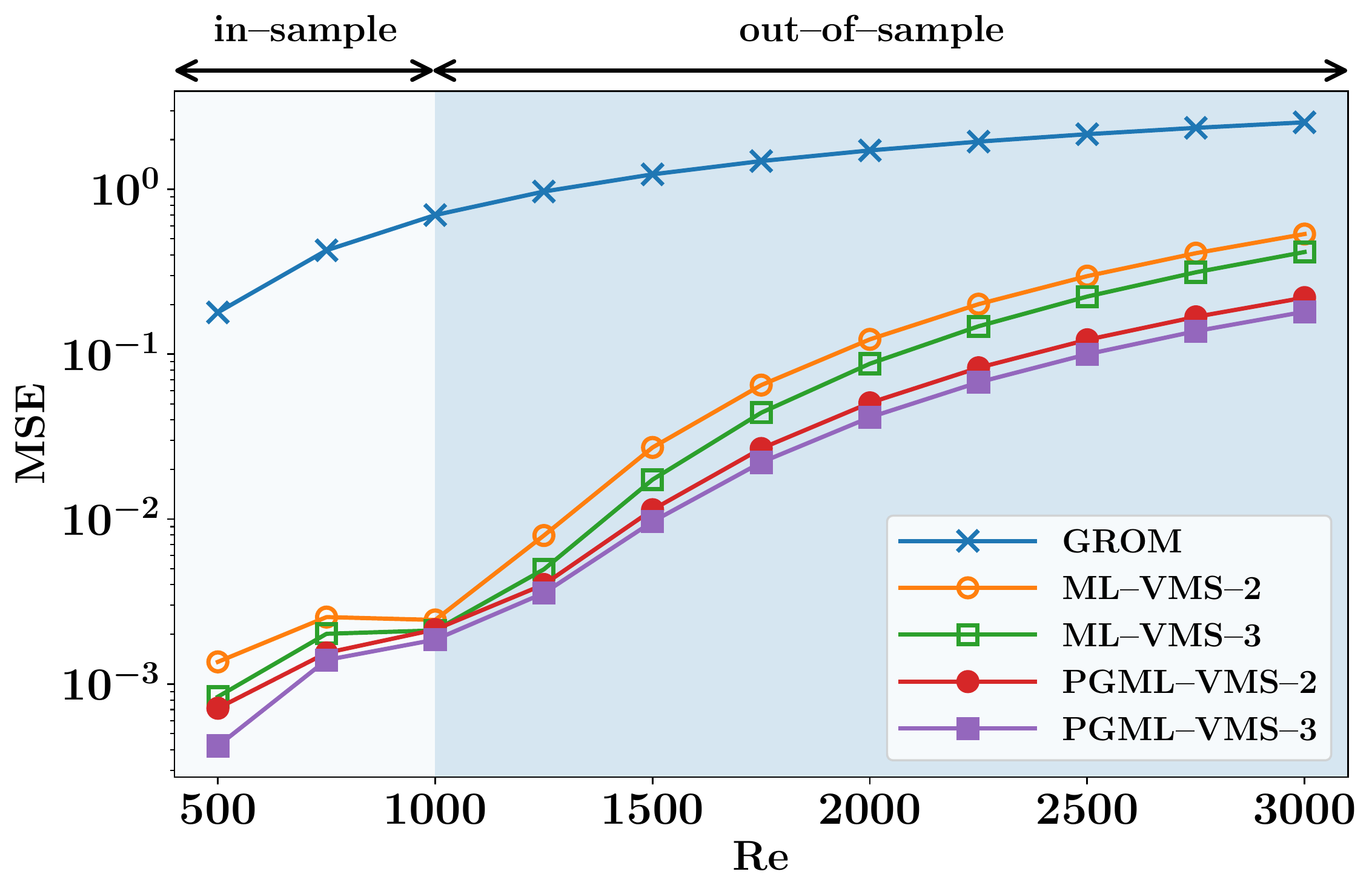}
    \caption{Mean squared error (MSE) between the true values of modal coefficients and the predictions of GROM, ML-VMS-2, ML-VMS-3, PGML-VMS-2, and PGML-VMS-3.}
	\label{fig:mse}
\end{center}
\end{figure}

Finally, the PGML-VMS-3 results are shown in~\cref{fig:pgml3}, where we can see improved results across all the resolved scales with very low levels of uncertainty. The mean squared error (MSE) between the true projection values of the resolved scales and the prediction of the %GROM 
ROM with and without various closure models is shown in~\cref{fig:mse}. We can see that the VMS closure provides at least one order of magnitude better predictions than the baseline GROM. Moreover, the PGML-VMS is superior to the ML-VMS, especially for Reynolds number values that are not included in the LSTM training. This can be attributed to the fact that PGML employs physics-based features derived from the governing equations, resulting in improved extrapolatory capabilities of the overall model. Finally, the three-level variant of VMS is providing more accurate ROMs than VMS-2, making use of the locality of information %energy 
transfer to build more localized closure models.

\clearpage

\subsection{NLPOD for unresolved scales} \label{sec:6.2}

\begin{figure}[h]
\begin{center}
	\includegraphics[width=0.75\textwidth]{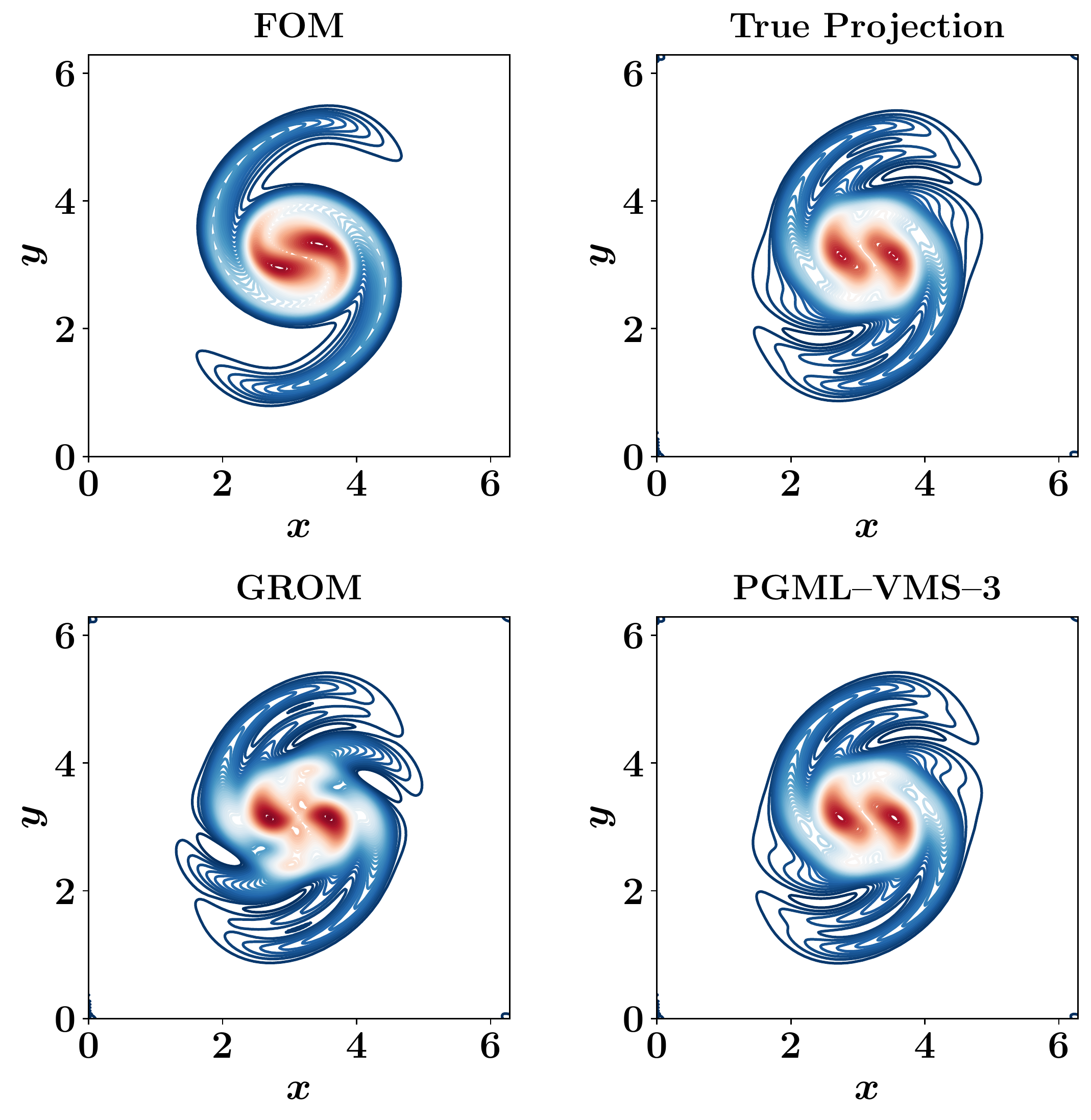}
    \caption{Comparison between the FOM vorticity field at the final time (i.e., $t=30$) %against 
    and the reconstruction from true projection (i.e., optimal reconstruction), GROM, and PGML-VMS-3. Note that the PGML-VMS-3 field is very similar to the true projection field, which implies that the closure error is minimized. %Despite that, there is a clear discrepancy, compared to the FOM results, which indicates a significant projection error.
    However, there are clear differences between the FOM and PGML-VMS-3 results, which suggest a significant projection error in the PGML-VMS-3 model.
    }
	\label{fig:field}
\end{center}
\end{figure}

\begin{figure}[h]
\begin{center}
	\includegraphics[width=0.75\textwidth]{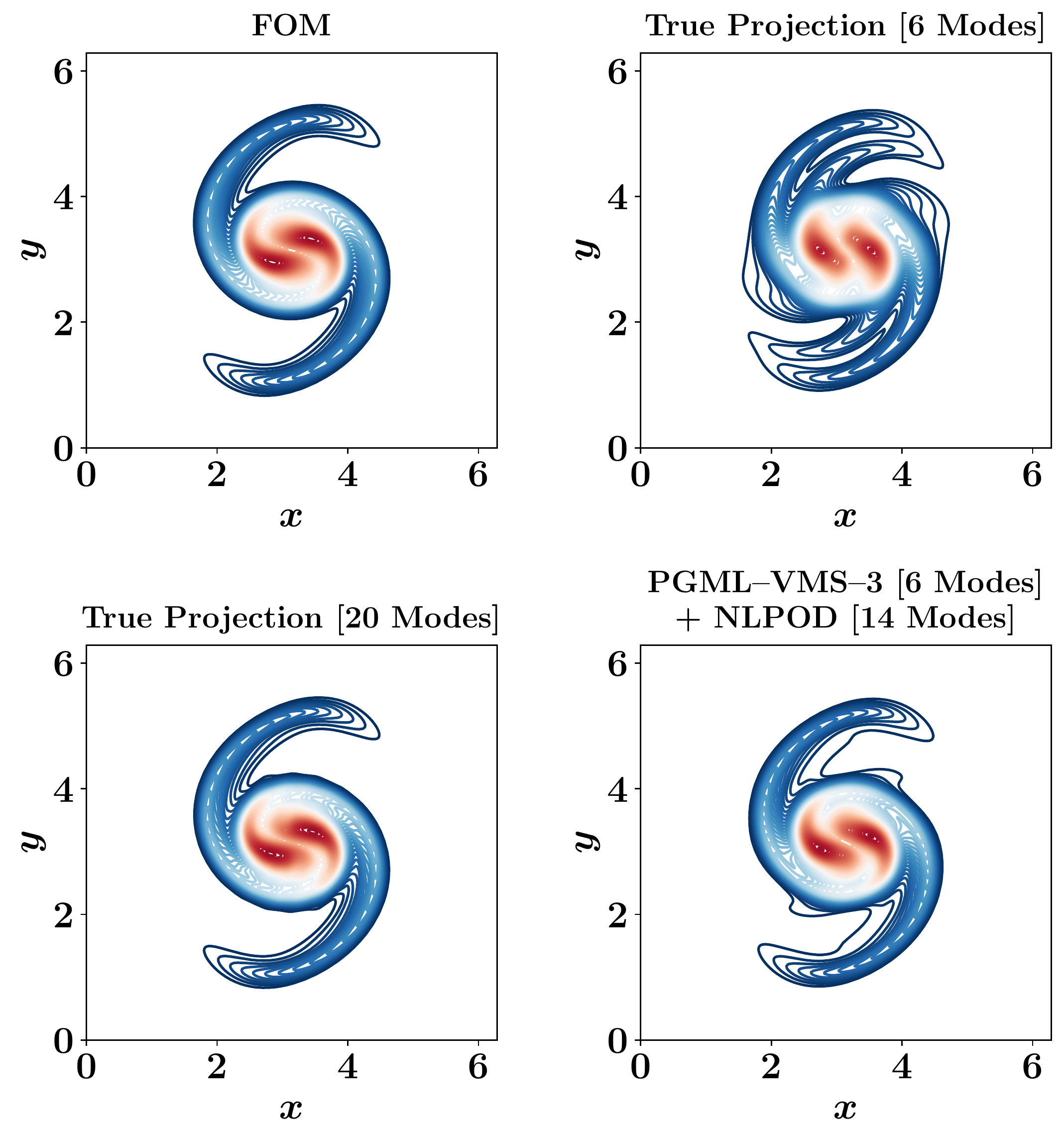}
    \caption{Comparison between the FOM vorticity field at final time (i.e., $t=30$) %against 
    and the reconstruction from true projection (i.e., optimal reconstruction) at two different values of modal truncation, as well as the predictions of the PGML-VMS-3 for the dynamics of the first 6 modes, augmented with NLPOD for the following 14 modes (i.e., a total of $K=20$ modes) to reduce the projection error.}
	\label{fig:nlpod}
\end{center}
\end{figure}

The reconstructed vorticity fields from GROM, true projection, and PGML-VMS-3 at final time (i.e., $t=30$) at $\text{Re}=3000$ are visualized in~\cref{fig:field}. We can see that the GROM field is significantly inaccurate. In contrast, the PGML-VMS-3 is very close to the true projection field. This %implies 
suggests that the PGML-VMS-3 is successful in providing accurate closure terms in such a way that the resulting ROM trajectory converges to the best linear approximation with 6 modes. Nonetheless, compared to the FOM solution, it is clear that 6 POD modes are not enough to capture all the relevant flow structures, especially at large values of the Reynolds number. On the other hand, building a projection-based ROM with increased number of modes will result in an undesired higher computational %burdens. 
burden. 
In order to cure this limitation, we apply the NLPOD methodology from~\cref{sec:nlpod} to learn a latent space representation of important unresolved scales. We find that the value $K=20$ corresponds to $\text{RIC}\ge 99.99\%$, so we consider $\bb = \{a_k\}_{k=7}^{20} \in \mathbb{R}^{14}$ in the NLPOD extension. We use the NLPOD to learn a two-dimensional compression of the resolved scales, i.e., $\bz = \{z_k\}_{k=1}^{2}\in \mathbb{R}^{2}$. \cref{fig:nlpod} displays the reconstructed vorticity fields at the final time from the true projection of the FOM field onto the first 6 and the first 20 POD modes. We notice that the FOM flow scales can be adequately captured by the subspace spanned by the first 20 POD modes. Furthermore, %we 
the plots clearly 
show that the combination of PGML-VMS-3 for the first 6 modes and NLPOD for the subsequent 14 modes (i.e., a total of 20 modes) provides improved field reconstruction. We highlight that the computational overhead of the online deployment of the PGML-VMS closure and NLPOD is negligible compared to solving the projection-based ROM with 6 modes.
%\av{SORRY, I DO NOT GET THE PREVIOUS SENTENCE: WHAT ARE WE COMPARING? IF IT IS NLPOD vs GROM IT DOES NOT SEEM TO ME IT IS NEGLIGIBLE, BUT MAYBE I MISUNDERSTOOD}

The %computational costs in terms of 
CPU times for different portions of the %considered frameworks for 
FOM and ROMs are listed in~\cref{table:CPU}. For the ROMs, we can see that the majority of the time is spent to train the neural networks during the offline stage. We note that this time can be %reduced 
significantly reduced by considering parallel training algorithms that make use of distributed hardware facilities. We also notice that the three-level VMS framework takes about twice the time taken by the two-level VMS due to the use of two distinct neural networks, which doubles the training and testing time. Nonetheless, we see that considerable computational gains are achieved compared to the FOM, by offloading most of the expensive computations to the offline stage resulting in computationally light models that can be used efficiently in the online stage. Moreover, we notice that the costs of the ML and PGML frameworks are of the same order, which implies that incorporating physics-based features into the neural network latent space comes with negligible overheads. 

\begin{table*}[htbp!]
\caption{Comparison of the CPU times for the offline and online stages for FOM and ROMs. Note that the PGML-VMS-3+NLPOD model yields a level of accuracy which is similar to the GROM ($R=20$) model with only a fraction of computational overhead (i.e., with a total computational online execution time of 63.876 s for the PGML-VMS-3+NLPOD model).}
\centering
\begin{tabular}{p{0.35\textwidth} p{0.10\textwidth} || p{0.30\textwidth}  p{0.11\textwidth} }  
\hline
\multicolumn{2}{c}{Offline CPU Time [s]} & \multicolumn{2}{c}{Online CPU Time [s]} \\
\hline
POD Basis &  0.646  &  FOM & 1860.056 \\
GROM Operators  & 0.246 & GROM ($R=6$) & 20.226 \\
ML-VMS-2 Training & 71.641 & ML-VMS-2 ($R=6$) & 32.289  \\
ML-VMS-3 Training & 148.057 & ML-VMS-3 ($R=6$) & 45.055 \\
PGML-VMS-2 Training & 65.324 & PGML-VMS-2 ($R=6$) & 33.358 \\
PGML-VMS-3 Training & 139.863 & PGML-VMS-3 ($R=6$) & 51.545 \\
NLPOD Training (AE) & 111.543 &  NLPOD ($R=6, K=20$) &  12.331 \\
NLPOD Training (LSTM) & 85.234 &  GROM ($R=20$) &  604.427\\ 
\hline \smallskip 
\end{tabular}
\label{table:CPU}
\end{table*}

%%=====================================%%
\section{Conclusions and Future Work} %Works} 
\label{sec:conc}
%%=====================================%%
We propose a hybrid hierarchical learning approach for the reduced order modeling of nonlinear fluid flow systems. The core component of the proposed method comprises %of 
a multi-level variational multiscale (VMS) framework for the %distinction between 
natural separation of 
the resolved modes of different length scales and unresolved modes. We develop a modular physics-guided machine learning (PGML) paradigm through the concatenation of neural network layers to enable the convergence of the ROM trajectory of resolved scales to the optimal low-rank approximation. We use the projection of the %universal 
governing equations onto the %respective 
POD modes as physics-based features to constrain the output to a manifold of the physically realizable solutions. 
For a vorticity transport problem with high Reynolds numbers, we numerically demonstrate that this injection of physical information yields more robust and reliable ROM closures with reduced uncertainty levels. %for the ROM of the vortex merger problem. 
Moreover, we showcase the benefits of exploiting the locality of information transfer by building a three-level VMS, which centers around the scale-separation of the resolved modes into large resolved scales and small resolved scales. %Results 
The numerical results 
show that the VMS-3 provides significant flexibility in defining the closure terms and is superior to the classical VMS-2 model used in previous studies. Finally, to decrease the projection error, we adapt the nonlinear proper orthogonal decomposition approach to learn a latent space representation of the unresolved ROM scales that yield a near-full rank approximation of the flow field.

Further investigations are required to optimize the layer(s) at which physics-based features are injected in the PGML framework. %s. 
For example, we can add the injection at multiple points in the latent space, rather than a single point. Moreover, we may fuse various information from different models by repeating the concatenation operator for each piece of information. It is worth noting that advanced hyperparameter tuning approaches for the automated design of neural network architectures (e.g., using genetic algorithms) can be utilized to find the optimal layer(s) to inject the physics in the PGML architectures. In the present study, the ML-VMS, PGML-VMS, and NLPOD components of the hybrid framework are treated separately. In other words, the training of each neural network takes place independently of other neural networks in the framework. In a follow-up study, we plan to explore the simultaneous training of these neural networks to ensure that these models are integrated seamlessly in the computational workflow. Finally, the truncated scales that are recovered by NLPOD can be further embedded in the PGML-VMS architecture to improve the approximation of the closure model.

%This is indicated for high Reynolds number problems.

% \AV{Is the following list a ``roadmap'' for a follow-up or something already implemented? If it is a follow-up, maybe we can write this in the Conclusion section.}
% \ti{I agree.  I think we should say something in the conclusions.}

\section*{Acknowledgments}
%\AV{It seems that the acknowledgements are duplicated.}
%This material is based upon work supported by the U.S. Department of Energy, Office of Science, Office of Advanced Scientific Computing Research under Award Number DE-SC0019290. O.S. gratefully acknowledges the Early Career Research Program (ECRP) support of the U.S. Department of Energy. O.S. also gratefully acknowledges the financial support of the National Science Foundation under Award Number DMS-2012255.
%O.S. gratefully acknowledges the financial support of the National Science Foundation under Award Number DMS-2012255 and the Early Career Research Program (ECRP) support of the U.S. Department of Energy under Award Number DE-SC0019290. 
%Research was supported by the NSF Project 2012254 (T.I.) - 2012255 (O.S.) - 2012286 (A.V.).
This research was supported by National Science Foundation grants DMS-2012253 (T.I.), DMS-2012255 (O.S.), and DMS-2012286 (A.V.).
O.S. also gratefully acknowledges support through the U.S. Department of Energy, Office of Science, Office of Advanced Scientific Computing Research under Award Number DE-SC0019290.
%T.I. acknowledges support through National Science Foundation Grant Number DMS-2012253.
%T.I. acknowledges support through National Science Foundation Grant Number DMS-2012253. 
%A.V. acknowledges support through National Science Foundation Grant Number .
A.V. also gratefully acknowledges the support of the 
National Science Foundation grant DMS-2038118.

\section*{Acknowledgements}
This research was supported by National Science Foundation grants DMS-2012253 (T.I.), DMS-2012255 (O.S.), and DMS-2012286 (A.V.).
O.S. also gratefully acknowledges support through the U.S. Department of Energy, Office of Science, Office of Advanced Scientific Computing Research under Award Number DE-SC0019290.
%T.I. acknowledges support through National Science Foundation Grant Number DMS-2012253.
%T.I. acknowledges support through National Science Foundation Grant Number DMS-2012253. 
%A.V. acknowledges support through National Science Foundation Grant Number .
A.V. also gratefully acknowledges the support of the 
National Science Foundation grant DMS-2038118.

%\section*{Data availability}

%\nolinenumbers
%\section*{References}
\bibliographystyle{unsrt} 

\bibliography{manuscript}

\end{document}